\documentclass[nofootinbib,aps,pra,twocolumn,superscriptaddress,floatfix]{revtex4-2}
\usepackage[english]{babel}
\usepackage[utf8]{inputenc}
\usepackage[T2A]{fontenc}
\usepackage{hyperref, xcolor, graphicx}
\usepackage{amsfonts, amssymb, amsmath, mathtools}
\usepackage{subcaption}
\usepackage{siunitx}
\usepackage{tikz}
\usepackage{calc}
\usepackage{caption}
\usepackage{ragged2e}
\usepackage{float}

\hypersetup{
    colorlinks=true,
    linkcolor=black,
    citecolor=blue,   
    urlcolor=blue,
}
\begin{document}

\title{Peculiarities of spin dynamics excitation by magnetic field of a high-frequency electromagnetic pulse}

\author{N.I.\ Gribova}
    \email{gribova.ni@phystech.edu}
    \affiliation{Lomonosov Moscow State University, Moscow 119991, Russia}
    \affiliation{Russian Quantum Center, Moscow 121205, Russia}
    \affiliation{Moscow Institute of Physics and Technology (National Research University), Dolgoprudny 141701, Russia}
\author{A.K.\ Zvezdin}
    \affiliation{Russian Quantum Center, Moscow 121205, Russia}
\author{V.I.\ Belotelov}
    \affiliation{Lomonosov Moscow State University, Moscow 119991, Russia}
    \affiliation{Russian Quantum Center, Moscow 121205, Russia}

\date{\today}

\begin{abstract}
Terahertz (THz) electromagnetic pulses offer a promising route for the ultrafast manipulation of magnetization in ferromagnetic materials. While previous studies have demonstrated the excitation of spin dynamics using linearly polarized THz fields, the role of circular polarization and the effects of rapidly oscillating, time-dependent field profiles remained insufficiently understood. We have developed a unified theoretical framework for describing the excitation of spin precession via Zeeman interaction in magnetic materials by high frequency pulses of arbitrary polarization with temporal Gaussian profile. 
In the regime of long pulses (at least several oscillations are within the pulse duration), a circularly polarized magnetic field acts as an effective rectified magnetic field along the pulse propagation, while linear polarized pulses excite no free precession. In the regime of short pulses (less than one oscillation is within the pulse duration), pulses of any polarization, including linear one can excite free spin precession. There is an optimal pulse duration which maximizes amplitude of the spin precession. It depends on magnetic parameters of the sample and the external magnetic field, as well as on the carrier frequency of the pulse and its amplitude. These findings bridge key gaps in the understanding of THz-induced spin dynamics and provide insights into the design of light-controlled magnetization schemes using tailored electromagnetic pulses.
\end{abstract}

\maketitle

\section{Introduction}

Terahertz (THz) electromagnetic waves have emerged as a powerful tool for driving ultrafast spin dynamics in magnetic materials through light-matter interaction. In particular, linearly polarized THz pulses can excite resonant spin oscillations in magnets via the Zeeman interaction, where the magnetic field component of the pulse couples directly to the spin system \cite{lu2017coherent, bocklage2016model}. 
Extensive experimental studies across multiple groups \cite{mikhaylovskiy2015terahertz,mikhaylovskiy2016colossal, blank2021thz, grishunin2023two} have demonstrated THz-driven spin dynamics in both ferromagnets \cite{shalaby2016simultaneous, vicario2013off, unikandanunni2022inertial} and antiferromagnets \cite{kampfrath2011coherent, jin2013single, baierl2016terahertz, mashkovich2019terahertz, grishunin2021excitation, blank2023empowering}.

Generally, spin dynamics is excited by single-cycle THz pulses in the magnetic materials whose spin resonances are at sub-THz frequencies, i.e. lie in the vicinity of the THz pulse carrier frequency and within the THz pulse spectrum. In antiferromagnets such as rare-earth orthoferrites and iron borate, single-cycle THz fields resonantly drive the quasi-antiferromagnetic modes whose eigenfrequencies lie in the sub-THz/THz range, enabling phase-coherent precession of the Néel vector \cite{jin2013single,mashkovich2019terahertz,blank2023empowering}. In FeBO\textsubscript{3}, for example, intense nearly single-cycle THz pulses resonantly access the high-frequency quasi-antiferromagnetic mode ($\sim$0.5$\,$THz) \cite{mashkovich2019terahertz,blank2023empowering}. Similar resonant modes in orthoferrites (e.g., NdFeO\textsubscript{3}, ErFeO\textsubscript{3}) appear around 0.3–0.5$\,$THz and are efficiently excited and/or probed when the resonance lies within the spectrum of the incident THz pulse.

Nevertheless, THz pulses can also be used to control spin dynamics off-resonantly in materials with GHz spin resonances \cite{kampfrath2013resonant, vicario2013off, bocklage2016model,vicario2014terahertz,rongione2023emission, hudl2019nonlinear, unikandanunni2022inertial}. Thus, it was demonstrated in \cite{vicario2013off} that short THz pulses drive a coherent subcycle femtosecond magnetization dynamics in ferromagnetic thin films. The spin dynamics excited by such method is a kind of forced precession and exists only within the THz pulse duration. Bocklage in \cite{bocklage2016model} developed a theory describing such experiments and extended them by modeling the effect of single- and multiple-cycle THz pulses. It was shown that both the shape of the pulse and its polarization can be used to control the magnetization trajectories. Ref. \cite{shalaby2017off} investigated off‑resonant magnetization dynamics in Co, Fe и Ni thin films driven by intense single‑cycle THz fields, confirming and expanding the Vicario results  \cite{vicario2013off,vicario2014terahertz}. More recently, Ref. \cite{rongione2023emission} reported emission of coherent THz magnons in an antiferromagnetic insulator via ultrafast spin–phonon interactions — another manifestation of the off‑resonant ultrafast spin dynamics.

Despite these advances, the problem of the off-resonant spin excitation still requires further studies. By varying THz pulse duration and polarization it might be possible to achieve different scenarios of the lower frequency spin dynamics which is quite important for spin control and manipulation in magnets with ferromagnetic modes. 

To make further progress in this direction we present a unified theoretical framework for the description of the off-resonant excitation of magnetic oscillations by pulses of fast oscillating magnetic field with arbitrary polarization and time-dependent profiles. We investigate the excitation of spin dynamics in a sample subjected to short and long laser pulses with different polarizations. For short pulses, we identify the conditions under which the maximum amplitude of the dynamical response is achieved. In the case of long pulses, we find that spin dynamics are excited exclusively under circular polarization, while the linear polarization does not induce any free spin precession. Section II presents a general theory of spin dynamics under excitation by the Gaussian pulses of arbitrary polarization. Section III reports numerical simulations for both long and short Gaussian pulses of arbitrary polarization, and provides a description of the excitation process in terms of the effective magnetic field.

\section{Theory}

We consider linearized equations of spin dynamics near the ground state of a magnetic film with uniaxial magnetic anisotropy in the external magnetic field $H_{ext}$ along $y$ axis (Fig. \ref{fig:scheme1}) (Supplementary, Section A). Let's assume that the sample is subjected to a high frequency electromagnetic field of an arbitrary polarization. We will consider action of its magnetic field component on spins. The oscillating magnetic field is represented by $\mathbf{h} = \frac{1}{2}(\mathbf{h}_0 e^{i\omega t+i\varphi} +\mathbf{h}_0^{*} e^{-i\omega t-i\varphi})f(t)$, where $\omega$ is frequency, $\varphi$ is the initial phase, $\tau$ is characteristic duration of the pulse, and $f(t)$ is its envelope function.

\setcounter{figure}{0}
 \begin{figure}[h]
    \centering
    \tikzset{every picture/.style={line width=0.75pt}} 

\begin{tikzpicture}[x=0.75pt,y=0.75pt,yscale=-0.5,xscale=0.5]

\draw   (113.48,82.01) -- (338.83,82.01) -- (242.25,136.17) -- (16.9,136.17) -- cycle ;
\draw   (15.82,136.28) -- (241.94,136.28) -- (241.94,168.67) -- (15.82,168.67) -- cycle ;
\draw    (339.86,81.68) -- (339.86,114.07) ;
\draw    (241.94,168.67) -- (340.81,114.07) ;
\draw    (186.12,174.52) .. controls (186.12,180.26) and (185.67,184.27) .. (186.88,187.71) ;
\draw   (179.61,183.67) -- (186.43,174.21) -- (193.26,183.67) ;
\draw    (115.91,88.16) -- (116.22,32.25) ;
\draw [shift={(116.23,30.25)}, rotate = 90.32] [color={rgb, 255:red, 0; green, 0; blue, 0 }  ][line width=0.75]    (10.93,-3.29) .. controls (6.95,-1.4) and (3.31,-0.3) .. (0,0) .. controls (3.31,0.3) and (6.95,1.4) .. (10.93,3.29)   ;
\draw    (115.91,88.16) -- (178.59,88.01) ;
\draw [shift={(180.59,88.01)}, rotate = 179.86] [color={rgb, 255:red, 0; green, 0; blue, 0 }  ][line width=0.75]    (10.93,-3.29) .. controls (6.95,-1.4) and (3.31,-0.3) .. (0,0) .. controls (3.31,0.3) and (6.95,1.4) .. (10.93,3.29)   ;
\draw    (115.91,88.16) -- (76.87,111.04) ;
\draw [shift={(75.15,112.05)}, rotate = 329.63] [color={rgb, 255:red, 0; green, 0; blue, 0 }  ][line width=0.75]    (10.93,-3.29) .. controls (6.95,-1.4) and (3.31,-0.3) .. (0,0) .. controls (3.31,0.3) and (6.95,1.4) .. (10.93,3.29)   ;
\draw  [color={rgb, 255:red, 155; green, 155; blue, 155 }  ,draw opacity=1 ] (140.15,118.23) .. controls (139.02,110.44) and (159.22,100.94) .. (185.25,97.02) .. controls (211.29,93.09) and (233.3,96.22) .. (234.43,104) .. controls (235.55,111.79) and (215.35,121.29) .. (189.32,125.22) .. controls (163.28,129.14) and (141.27,126.02) .. (140.15,118.23) -- cycle ;
\draw   (149.74,120.43) -- (156.78,126.53) -- (146.95,127.69) ;
\draw    (57.76,152.26) -- (198,152.68) ;
\draw [shift={(200,152.69)}, rotate = 180.17] [color={rgb, 255:red, 0; green, 0; blue, 0 }  ][line width=0.75]    (10.93,-3.29) .. controls (6.95,-1.4) and (3.31,-0.3) .. (0,0) .. controls (3.31,0.3) and (6.95,1.4) .. (10.93,3.29)   ;
\draw  (367.7,150.51) -- (630.71,150.51)(498.33,43.09) -- (498.33,253.74) (623.71,145.51) -- (630.71,150.51) -- (623.71,155.51) (493.33,50.09) -- (498.33,43.09) -- (503.33,50.09)  ;
\draw  [color={rgb, 255:red, 155; green, 155; blue, 155 }  ,draw opacity=1 ] (397.3,213.44) .. controls (385.04,192.57) and (420.33,147.47) .. (476.12,112.71) .. controls (531.92,77.96) and (587.09,66.71) .. (599.35,87.58) .. controls (611.61,108.46) and (576.32,153.55) .. (520.53,188.31) .. controls (464.74,223.07) and (409.57,234.32) .. (397.3,213.44) -- cycle ;
\draw  [dash pattern={on 4.5pt off 4.5pt}]  (366.11,235.42) -- (623.82,70.94) ;
\draw  [draw opacity=0] (497.61,133.04) .. controls (504.08,132.84) and (509.7,135.17) .. (512.34,139.76) .. controls (512.59,140.2) and (512.81,140.66) .. (513,141.12) -- (492.51,150.31) -- cycle ; \draw   (497.61,133.04) .. controls (504.08,132.84) and (509.7,135.17) .. (512.34,139.76) .. controls (512.59,140.2) and (512.81,140.66) .. (513,141.12) ;  
\draw   (486.38,153.46) -- (497.89,151.26) -- (491.51,161.37) ;
\draw    (400.17,213.7) -- (494.97,153.18) ;
\draw    (476.12,112.71) -- (498.33,150.51) ;
\draw   (479.08,125.85) -- (476.45,113.9) -- (486.39,120.72) ;
\draw   (410.43,212.4) -- (398.93,214.62) -- (405.28,204.5) ;
\draw   (493.76,139.24) -- (498.89,150.58) -- (487.63,145.33) ;
\draw [color={rgb, 255:red, 208; green, 2; blue, 27 }  ,draw opacity=1 ]   (185.26,111.52) -- (232.45,104.31) ;
\draw [shift={(234.43,104)}, rotate = 171.3] [color={rgb, 255:red, 208; green, 2; blue, 27 }  ,draw opacity=1 ][line width=0.75]    (8.74,-2.63) .. controls (5.56,-1.12) and (2.65,-0.24) .. (0,0) .. controls (2.65,0.24) and (5.56,1.12) .. (8.74,2.63)   ;
\draw    (183.92,270) .. controls (182.95,262.67) and (183.92,261.47) .. (185.88,258.72) .. controls (188.82,255.26) and (212.3,249.41) .. (184.9,247.86) .. controls (162.39,247.86) and (171.2,253.29) .. (184.9,253.29) .. controls (193.71,253.71) and (242.64,240.88) .. (184.9,237.78) .. controls (149.67,237.78) and (148.69,242.43) .. (184.9,243.21) .. controls (227.96,242.43) and (250.14,228.09) .. (185.88,226.92) .. controls (119.33,226.14) and (145.76,234.67) .. (185.88,232.35) .. controls (223.37,232.35) and (271.02,219.94) .. (185.88,216.06) .. controls (130.1,214.51) and (121.62,221.49) .. (185.88,221.49) .. controls (244.79,220.32) and (238.45,207.15) .. (184.9,205.98) .. controls (138.35,205.93) and (141.84,210.63) .. (184.9,211.41) .. controls (246.56,209.08) and (204.48,197.44) .. (184.9,197.44) .. controls (155.54,197.44) and (171.2,202.1) .. (183.92,202.1) .. controls (220.13,201.32) and (187.72,192.62) .. (186.88,187.71) ;
\draw [color={rgb, 255:red, 155; green, 155; blue, 155 }  ,draw opacity=1 ] [dash pattern={on 4.5pt off 4.5pt}]  (139.93,112.8) -- (139.93,231.07) ;
\draw [color={rgb, 255:red, 155; green, 155; blue, 155 }  ,draw opacity=1 ] [dash pattern={on 4.5pt off 4.5pt}]  (234.43,94.24) -- (234.43,243.25) ;
\draw [color={rgb, 255:red, 208; green, 2; blue, 27 }  ,draw opacity=1 ]   (187.42,213.39) -- (225,212.78) ;
\draw [shift={(227,212.75)}, rotate = 179.07] [color={rgb, 255:red, 208; green, 2; blue, 27 }  ,draw opacity=1 ][line width=0.75]    (8.74,-2.63) .. controls (5.56,-1.12) and (2.65,-0.24) .. (0,0) .. controls (2.65,0.24) and (5.56,1.12) .. (8.74,2.63)   ;
\draw    (199.99,70.24) -- (309.74,69.84) ;
\draw [shift={(311.74,69.83)}, rotate = 179.79] [color={rgb, 255:red, 0; green, 0; blue, 0 }  ][line width=0.75]    (10.93,-3.29) .. controls (6.95,-1.4) and (3.31,-0.3) .. (0,0) .. controls (3.31,0.3) and (6.95,1.4) .. (10.93,3.29)   ;

\draw (58.01,113.05) node [anchor=north west][inner sep=0.75pt]  [font=\footnotesize] [align=left] {\textit{{\fontfamily{ptm}\selectfont {\large x}}}};
\draw (169.88,58.4) node [anchor=north west][inner sep=0.75pt]  [font=\footnotesize] [align=left] {{\fontfamily{ptm}\selectfont {\large y}}};
\draw (94.82,18.83) node [anchor=north west][inner sep=0.75pt]  [font=\footnotesize] [align=left] {{\large {\fontfamily{ptm}\selectfont z}}};
\draw (13.65,21.4) node [anchor=north west][inner sep=0.75pt]  [font=\footnotesize] [align=left] {{\fontfamily{ptm}\selectfont {\Large a)}}};
\draw (503.73,103.55) node [anchor=north west][inner sep=0.75pt]  [font=\footnotesize] [align=left] {$\displaystyle \mathrm{\psi }$};
\draw (630.94,125.56) node [anchor=north west][inner sep=0.75pt]  [font=\footnotesize] [align=left] {\textit{{\fontfamily{ptm}\selectfont {\large x}}}};
\draw (475.42,29.17) node [anchor=north west][inner sep=0.75pt]  [font=\footnotesize] [align=left] {{\fontfamily{ptm}\selectfont {\large y}}};
\draw (340.9,156.67) node [anchor=north west][inner sep=0.75pt]  [font=\scriptsize,rotate=-359.57] [align=left] {$\displaystyle \sqrt{1-\alpha ^{2}} h_{0}$};
\draw (440.65,123.83) node [anchor=north west][inner sep=0.75pt]  [font=\scriptsize] [align=left] {$\displaystyle \alpha h_{0}$};
\draw (311.49,49.86) node [anchor=north west][inner sep=0.75pt]  [font=\footnotesize] [align=left] {$\displaystyle \mathbf{H}_{\mathrm{ext}}$};
\draw (376.28,25.21) node [anchor=north west][inner sep=0.75pt]  [font=\footnotesize] [align=left] {{\fontfamily{ptm}\selectfont {\Large b)}}};
\draw (210.96,142.04) node [anchor=north west][inner sep=0.75pt]  [font=\footnotesize] [align=left] {$\displaystyle \mathbf{M}$};
\draw (244.09,91.93) node [anchor=north west][inner sep=0.75pt]  [font=\footnotesize] [align=left] {$\displaystyle \mathbf{h}$};

\end{tikzpicture}
    \caption{\justifying{(a) Configuration of a magnetic sample with respect to the incident electromagnetic pulse. (b) Polarization state of the electromagnetic pulse with respect to its parameters $\alpha$ and $\Psi$.}}
    \label{fig:scheme1}
\end{figure}
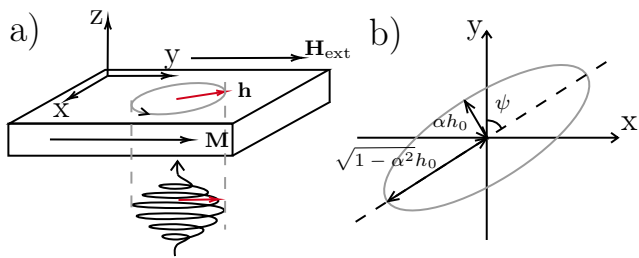
\setcounter{figure}{2}

We assume that the electromagnetic wave propagates along the normal of the magnetic film and, therefore, $h_z=0$, while the other two magnetic field components are given by $h_x = h_0 (\alpha\cos\psi\cos(\omega t+\varphi)+\sqrt{1-\alpha^2}\sin\psi\sin(\omega t+\varphi))f(t)$ and $h_y = h_0(-\alpha\sin\psi\cos(\omega t+\varphi)+\sqrt{1-\alpha^2}\cos\psi\sin(\omega t+\varphi))f(t)$, where normalization on the electromagnetic field energy is taken. Thus, the polarization of light is defined by two parameters $\alpha$  and $\psi$: $\alpha$ determines polarization of the electromagnetic pulse and takes values from zero to one, linear polarization corresponds to $\alpha=0, 1$, circular polarization --- to $\alpha=1/\sqrt2$, while $\psi$ is the angle of the main axis of the polarization ellipse to the equilibrium position of the magnetization. For example, if $\alpha=0$, then the angle $\psi=0$ represents linear polarization along $y$ axis, while $\psi=\pi/2$ describes linear polarization along $x$ axis. 

To obtain the equations of motion of the magnetization, we use Lagrangian formalism. The Lagrangian of a monodomain magnet film in a spherical coordinate system with polar angle $\theta$ and azimuthal angle $\phi$ (Fig. \ref{fig:scheme1}) is given by 
\begin{equation}
    \mathcal{L} = -\frac{M}{\gamma}\dot{\phi}\cos{\theta} -U_a-U_z -U_d,
\end{equation}
where $\gamma$ is gyromagnetic ratio, {$\textbf{M} = (M_x, M_y, M_z)= M(\sin\theta\cos\phi, \sin\theta\sin\phi, \cos\theta)$}, $M$ is saturation magnetization of the magnetic film. The first term of the Lagrangian represents kinetic part, while the second and the third are for potential part. The anisotropy energy for the uniaxial anisotropy is $U_a = K_u\cos^2\theta$, where $K_u$ is the magnetic anisotropy constant. Demagnetization energy is determined by $U_d=2\pi M^2 \cos^2\theta$. Zeeman energy $U_z$ consists of two parts: the one related to the high frequency oscillations of electro-magnetic field $V_h = -M(h_x\sin\theta\cos\phi+h_y\sin\theta\sin\phi)$ and the other one due to the constant external magnetic field along $y$ axis $-M H_y\sin\theta\sin\phi$. 
\setcounter{figure}{1}
\begin{figure*}[htb!] 
    \centering
    \begin{minipage}[b]{0.5\linewidth}
        \centering
        \includegraphics[width=0.97\linewidth]{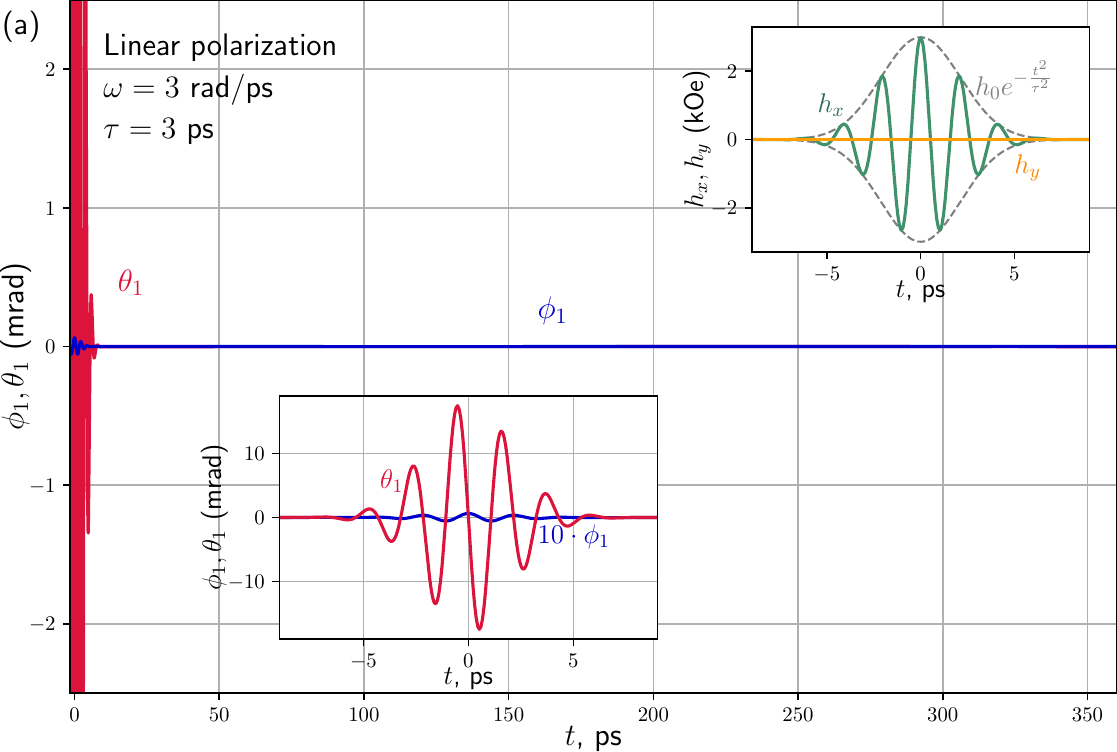}
    \end{minipage}%
    \begin{minipage}[b]{0.5\linewidth}
        \centering   \includegraphics[width=0.97\linewidth]{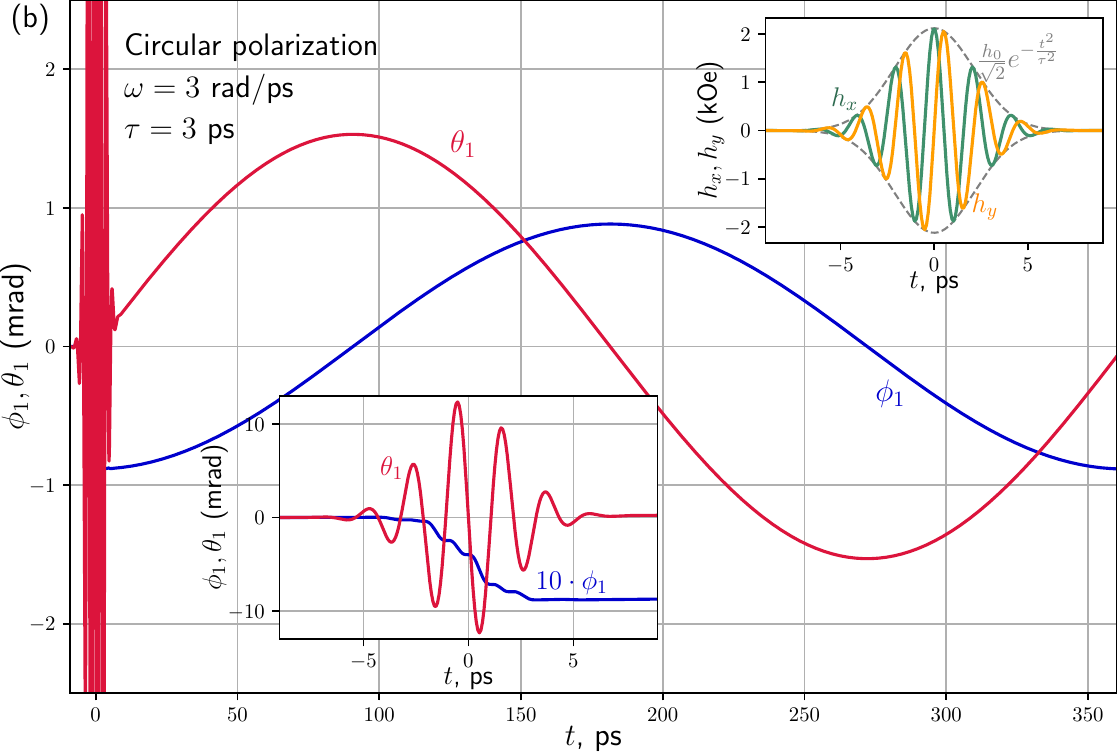}
    \end{minipage}
    \caption{\justifying{
    Spin dynamics excited by long electromagnetic pulses ($\tau\omega \gg 2\pi$) with linear (a) and circular (b) polarizations. Spin dynamics is calculated by Eqs. \eqref{phi1_dot} and \eqref{theta1_dot}. It is described by spherical-coordinate angles $\phi_1$ and $\theta_1$. A magnetic film with $\omega_r = 17.3~\mathrm{GHz}$ is exemplary considered. Electromagnetic pulse parameters are $\omega = 3~\mathrm{rad~ps^{-1}}$, $\tau = 3~\mathrm{ps}$, $h_0 = 3000~\mathrm{Oe}$, $\psi = 0$, and $\varphi = 0$.  Circular polarization refers to $\alpha = 1/\sqrt{2}$, linear polarization (right shown) --- to $\alpha = 1$. Bottom insets display initial part of the excitation process where forced oscillations are present. Top inserts represent $h_x$ and $h_y$ magnetic field components of the electromagnetic pulse. 
    }
    }
    \label{fig:figure2}
\end{figure*}

If $K_u<2\pi M^2$ then the ferromagnetic film has effective magnetic anisotropy of "easy plane" and the ground state is given by $\phi_0=\frac{\pi}{2}$, $\theta_0=\frac{\pi}{2}$ which means that equilibrium magnetization lies in-plane and is along $y$ axis. Therefore, the angles that determine the orientation of the magnetization can be represented as $\phi = \phi_0+\phi_1$ and $\theta = \theta_0+\theta_1$ with small deviations from the ground state $\theta_1, \phi_1 \ll 1$ (Supplementary A). Taking into account terms up to the first order in $\theta_1, \phi_1$ we derive the Euler-Lagrange linearized equations:
\begin{align}
\dot{\phi}_1-\omega_1 \theta_1 &= \gamma h_y \theta_1,\label{phi1_dot}\\
\dot{\theta}_1 + \omega_2 \phi_1&=-\gamma(h_x+h_y\phi_1).\label{theta1_dot}
\end{align}
where  $\omega_1 = \frac{2(K_u-2\pi M^2)\gamma}{M}+\gamma H_{ext}$ and $\omega_2 = \gamma H_{ext}$. Therefore, the system exhibits free oscillations at a resonant frequency $\omega_r = \sqrt{\omega_1\omega_2}$. 

In this article, we only consider the off-resonant case when $\omega_r\ll\omega$ which will allow to tackle the problem of excitation of GHz frequency spin precession by the magnetic field of high-frequency electromagnetic pulse in THz or infrared spectral range.
The case of a Gaussian pulse $f(t) = e^{-\frac{t^2}{\tau^2}}$ is presented in detail. In the next section spin dynamics excited by the "short" {($\tau\omega<2\pi$)} and "long" ({$\tau\omega\gg 2\pi$}) electromagnetic pulses of circular or linear polarizations is discussed.

\section{{Results and discussion}}
For the numerical simulations we consider a sample with parameters typical for the iron garnets. These materials have high quality ferromagnetic resonance (FMR) and are widely used for the ultrafast magnetism studies. In our analysis, we assume a resonant frequency $\omega_r = 17.3$~GHz ($\omega_1 = 10$~GHz and $\omega_2 = 30$~GHz), corresponding to a gyromagnetic ratio $\gamma = 1.76 \times 10^{-5}~\text{ps}^{-1}\text{Oe}^{-1}$, $4 \pi M = 1750$~Gs, $K_u = 4.1\times 10^{4}~$erg cm$^{-3}$, and the external magnetic field $H_{\mathrm{ext}}=1700~\text{Oe}$. 

We consider THz pulses with an electric field strength of 0.9~MV/cm whose magnetic field amplitude is $h_0 = 3000$~Oe. Such field strength corresponds to typical THz pulses used for excitation of spin dynamics \cite{fulop2020laser}, \cite{wu2018highly}.  

\begin{figure*}[htb!] 
    \centering
    \begin{minipage}[b]{0.49\linewidth}
        \centering        \includegraphics[width=0.97\linewidth]{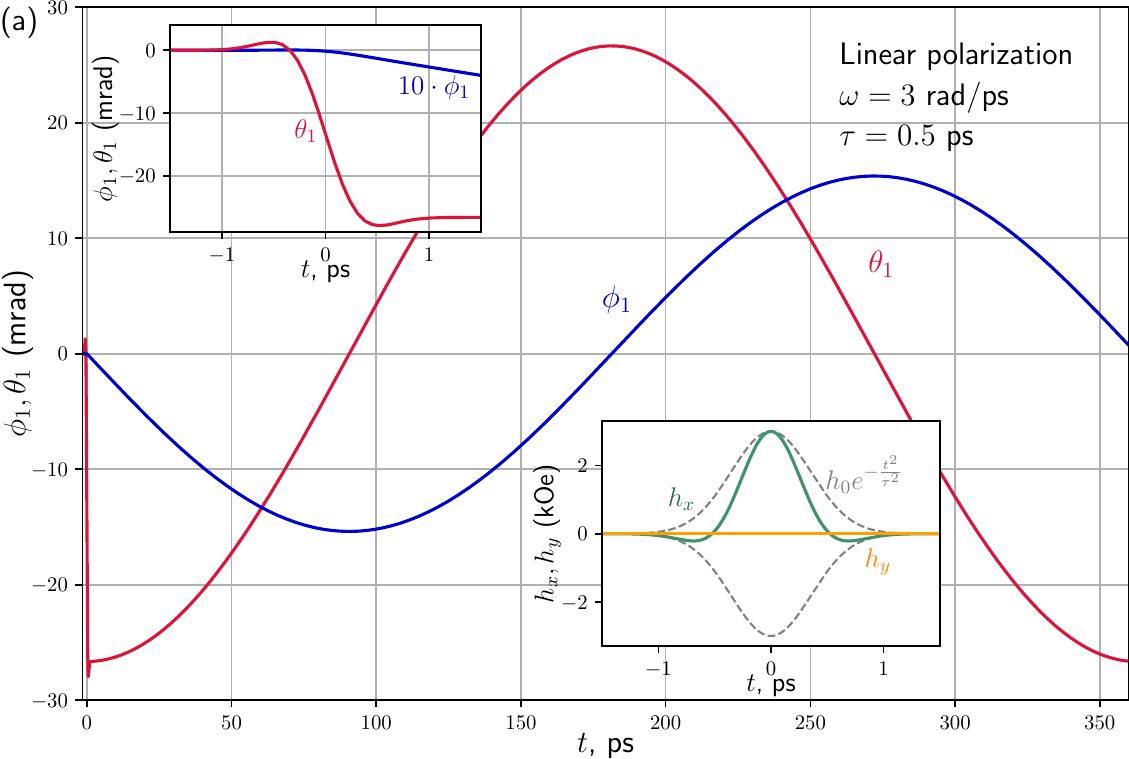}
    \end{minipage}
    \begin{minipage}[b]{0.49\linewidth}
        \centering        \includegraphics[width=0.97\linewidth]{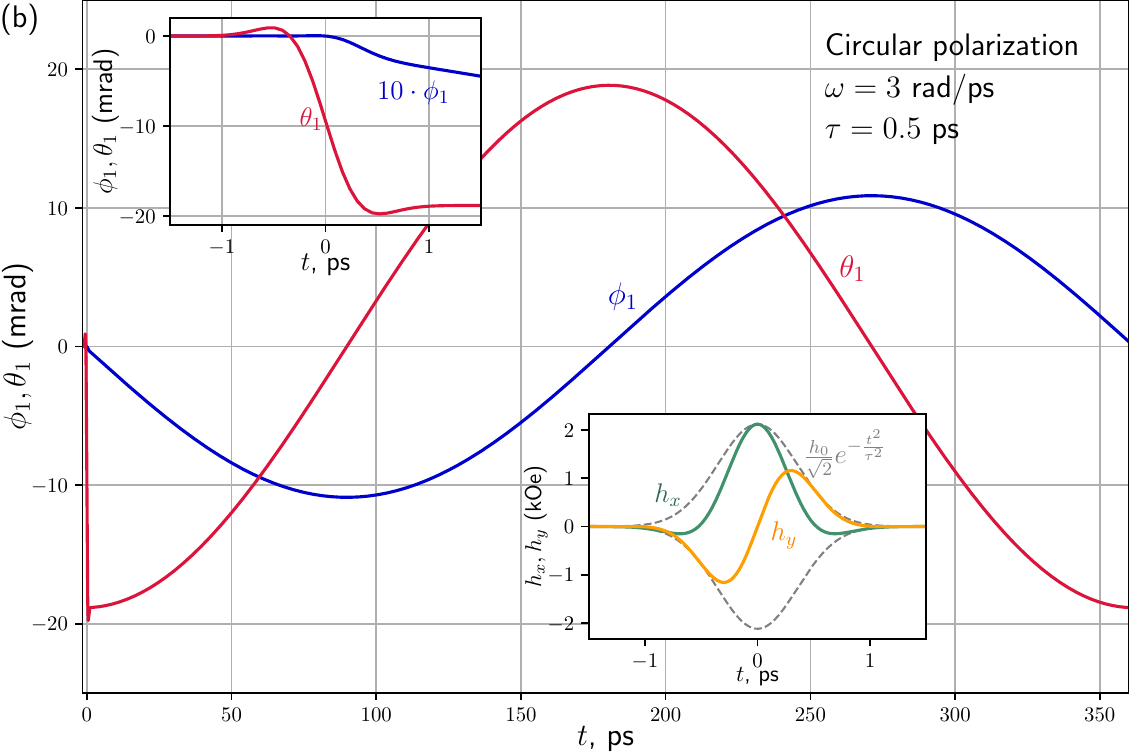}
    \end{minipage}
    \caption{\justifying
    {
    Spin dynamics excited by short electromagnetic pulses ($\tau\omega \gg 2\pi$) with linear (a) and circular (b) polarizations. Spin dynamics is calculated by Eqs. \eqref{phi1_dot} and \eqref{theta1_dot}. It is described by spherical-coordinate angles $\phi_1$ and $\theta_1$. A magnetic film with $\omega_r = 17.3~\mathrm{GHz}$ is exemplary considered. Electromagnetic pulse parameters are $\omega = 3~\mathrm{rad~ps^{-1}}$, $\tau = 0.5~\mathrm{ps}$, $h_0 = 3000~\mathrm{Oe}$, $\psi = 0$, and $\varphi = 0$.  Circular polarization refers to $\alpha = 1/\sqrt{2}$, linear polarization to $\alpha = 1$. Top insets display initial part of the excitation process where forced oscillations are present. Bottom inserts represent $h_x$ and $h_y$ magnetic field components of the electromagnetic pulse. 
    }
    }
    \label{fig:figure3}
\end{figure*}
\subsection{Long pulses}

We start from the case of long pulses $\tau \omega \gg 2\pi$, where the pulse contains a large number of electromagnetic wave periods.

The solutions of the linearized Eqs. \eqref{phi1_dot} and \eqref{theta1_dot} for an arbitrarily polarized Gaussian pulse are given by
\begin{align}
\phi_1(t\gtrsim 3\tau)&= -\frac{\beta \tau}{\omega}   \Biggl(
    c_1
    +c_2 e^{-\frac{\tau^2\omega_r^2}{2}}\Biggr)\cos\omega_rt 
    \label{phi_faraday_gauss},\\
\theta_1(t\gtrsim3\tau)&=\frac{\beta \tau}{\omega} \sqrt{\frac{\omega_2}{\omega_1}}
    \Biggl(c_1+c_2 e^{-\frac{\tau^2\omega_r^2}{2}}\Biggr)\sin\omega_rt.
    \label{theta_faraday_gauss}
\end{align}
where $\beta =\gamma^2h_0^2\frac{ \alpha \sqrt{1-\alpha^2}}{4}\sqrt{\frac{\pi}{2}}$, $c_1=1+\frac{\omega_1}{\omega_2}$, $c_2 = 1-\frac{\omega_1}{\omega_2}$. Equations \ref{phi_faraday_gauss} and \ref{theta_faraday_gauss} are valid when the excitation pulse is already finished, i.e. at the timescales $t\gtrsim 3\tau$. It should be noted that in this case spin dynamics is fully determined by the ellipticity parameter $\alpha$ and does not depend on the polarization angle $\psi$.

For linear polarization ($\alpha=1$ or $\alpha=0$, see top inset to Fig. 2(a)) $\beta$ in Eqs. \ref{phi_faraday_gauss} and \ref{theta_faraday_gauss} vanishes and no free oscillations remain after the Gaussian pulse ends (Fig. 2(a)). It is quite expected since fast oscillating magnetic field excites slowly oscillating spins ($\omega_r \ll \omega$). Nevertheless, at the timescales $t< 3\tau$ where Eqs. (4-5) are not valid, numerical solutions of Eqs. (1-2) gives forced oscillations at the electromagnetic wave frequency \(\omega\) at the timescales of the pulse duration (see bottom inset to Fig. 2(a)). 

However, if polarization of the pulse is circular, then situation drastically changes and the forced oscillations is followed by free oscillations taking place at the ns-timescales (Fig. 2(b)). Amplitude of the free precession increases as \(\tau\) grows (see inset to Fig. 4(a)). Similarly, elliptical polarization also gives birth to the long living precession, but the largest amplitude is achieved with circular polarization. Note, that for long pulses in both cases of linear and elliptical polarizations the spin dynamics is excited non-resonantly (Supplementary Section \ref{appendix_spectrum}).

Since the long electromagnetic pulses excite free spin precession this process can be considered in terms of an effective magnetic field acting at zero frequency on spins and directed along wavevector of the electromagnetic wave, i.e. along $z$ axis. As duration of the long pulse is still much shorter than the period of the free precession ($\tau \omega_r \ll 2\pi$) the process of its excitation can be considered as a kind of photonic strike on spins with an effective magnetic field (Supplementary Section \ref{appendix2}):  

\begin{equation}
    \mathbf{H}_{eff} =  -\frac{\gamma}{2\omega i}[\textbf{h}_0\times\textbf{h}_0^*] e^{-\frac{2 t^2}{\tau^2}}.
\end{equation}

The effective magnetic field is rectified, i.e. it follows envelope of the electromagnetic pulse and doesn't oscillate. If Eqs. \eqref{phi1_dot} and \eqref{theta1_dot} are solved with the effective field by Eq.(6) instead of fast oscillating fields $h(t)$ then they give spin dynamics identical to solutions of Eqs. \ref{phi_faraday_gauss}, \ref{theta_faraday_gauss}) for $t\gtrsim3\tau$, which confirms this reasoning. Therefore, the spin dynamics in the considered case can be calculated using the concept of the photonic strike, which significantly simplifies the analysis.

It should be noted that the effective magnetic field $\mathbf{H}_{eff}$ is consistent with the phenomenological description of the magnetodipole \cite{krinchik1969transparent} contribution to the inverse Faraday effect which appears due to a high frequency circularly polarized magnetic field of an electromagnetic wave influencing a magnetic sample \cite{gribova2024inverse,eleonskii1977effect} (Supplementary Section D). The magnetodipole inverse Faraday effect is a counterpart of the more famous electrodipole inverse Faraday effect \cite{pitaevskii1961electric, van1965optically, pershan1966theoretical}.

\subsection{Short pulses}
\begin{figure*}[htb!] 
    \centering
    \begin{minipage}[b]{0.5\linewidth}
        \centering
        \includegraphics[width=0.97\linewidth]{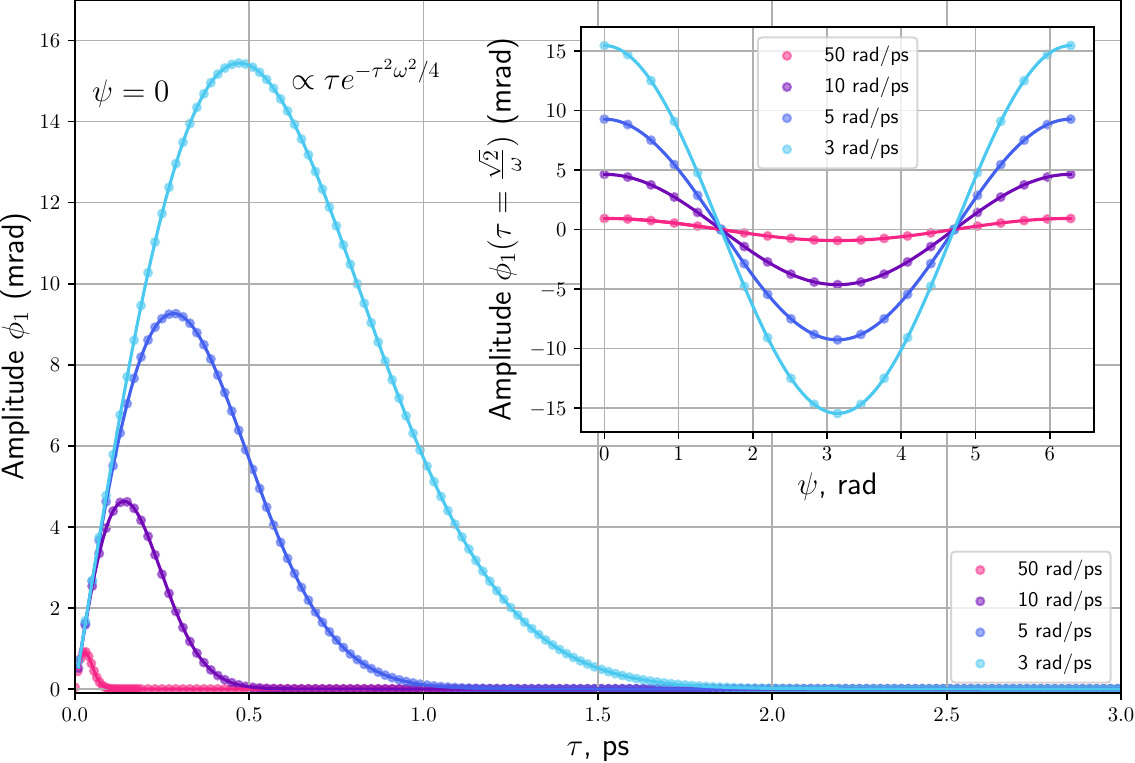}
        \subcaption{Linear polarization}
    \end{minipage}%
    \begin{minipage}[b]{0.5\linewidth}
        \centering
\includegraphics[width=0.97\linewidth]{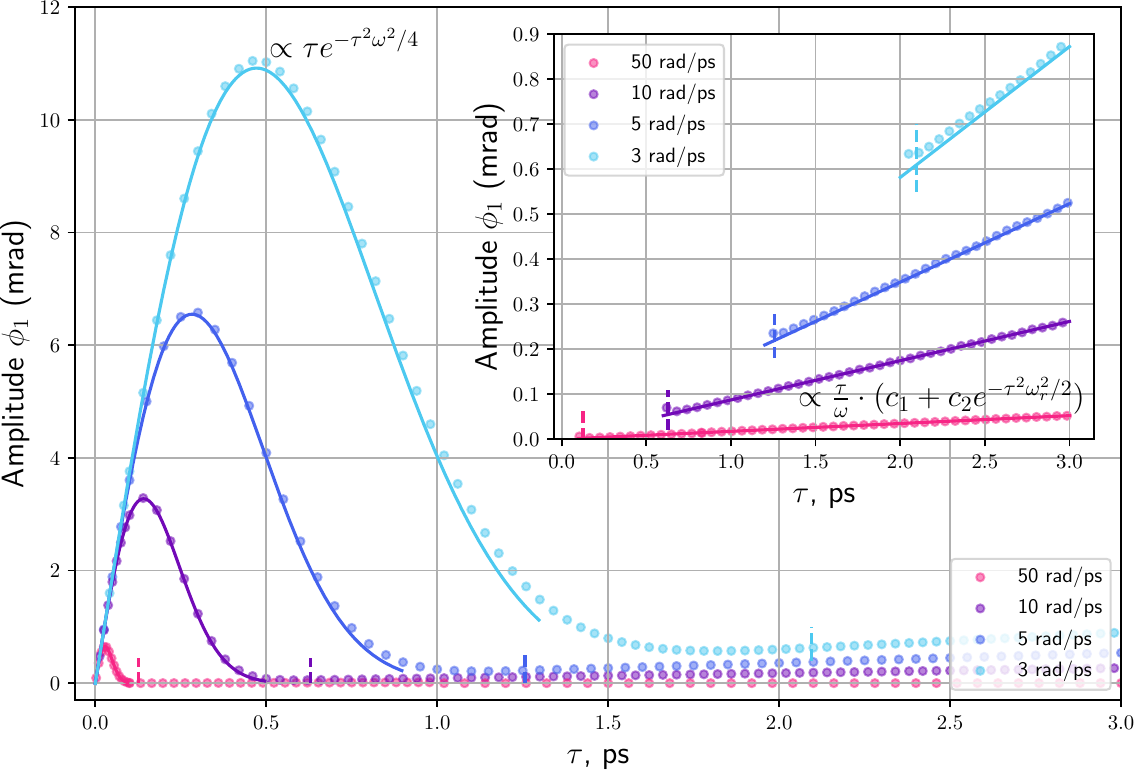}
        \subcaption{Circular polarization}
    \end{minipage}
    \caption{\justifying
    {
    Numerical and analytical solutions for the amplitude of spin precession \(\phi_1\) at the frequency \(\omega_r\). Dots represent numerical solutions of the linearized equations \eqref{phi1_dot} and \eqref{theta1_dot}, while solid lines correspond to analytical solutions.
    Panel (a) corresponds to linear polarization: solid lines represent the solutions of equations \eqref{theta_gauss_short_linear} and \eqref{phi_gauss_short_linear}, and the inset depicts the amplitude as a function of the angle \(\psi\) at \(\tau = \sqrt{2}/\omega\).  Panel (b) corresponds to circular polarization: solid lines given by equations \eqref{phi_gauss_short_linear} and \eqref{theta_gauss_short_linear} for \(\tau \omega \ll 2\pi\) are shown in the main figure, while those given by equations \eqref{phi_faraday_gauss} and \eqref{theta_faraday_gauss} for \(\tau \omega \gg 2\pi\) appear in the inset.
    The electromagnetic pulse parameters are \(\omega = 3, 5, 10, 50~\mathrm{rad}/\mathrm{ps}\), \(\tau = 0.5~\mathrm{ps}\), \(h_0 = 3000~\mathrm{Oe}\), and \(\varphi = 0\). Vertical dashed lines indicate \(\tau = 2\pi/\omega\).
    }
    }
    \label{fig:figure4}
\end{figure*}

In this section we consider the limit {$\tau\omega< 2\pi$}. Only the case of linear polarized short {G}aussian pulses is solved in the literature \cite{bocklage2016model} while here we propose a solution of the problem for an arbitrary polarization. The spin dynamics calculated by numerical solutions of Eqs. \ref{phi1_dot} and \ref{theta1_dot} in this case is presented in Fig. \ref{fig:figure3}.

In the limit $\tau\omega< 2\pi$ we get analytical solution in two cases. Unlike the case of long pulses, the short pulses induce a free precession not only for elliptical polarization but for the linear one as well: 

\begin{align}
\phi_1(t\gtrsim 3\tau)&= \gamma h_0\tau \zeta\cos\varphi \sqrt{\frac{\pi\omega_1}{\omega_2}}\,\,e^{-\tau^2\omega^2/4}\sin\omega_rt 
    \label{phi_gauss_short_linear},\\
\theta_1(t\gtrsim 3\tau)&=-\gamma h_0\tau \sqrt{\pi}\zeta\cos\varphi\,\, e^{-\tau^2\omega^2/4}\cos\omega_rt,
    \label{theta_gauss_short_linear}
\end{align}
\begin{equation}
    \mathbf{H}_{eff} = 2  h_0 \zeta \cos\psi\exp \Bigl[-\frac{\tau^2\omega^2}{4}-\frac{2t^2}{\tau^2}\Bigr]\,\mathbf{x},
\end{equation}

where $\zeta=\cos\psi$ for the linearly polarized pulse along $x$ axis ($\alpha=1$) and $\zeta=\alpha$ for the elliptically polarized pulse ($\psi=0$).
Therefore, in this case, similarly to the long pulses, one can also consider spin excitation as a kind of photonic strike by a non-oscillating effective magnetic field with time dependence following the pulse envelope. However, this time the effective field is directed along the $x$-axis (see Supplementary Section B).

In the case of linear polarization, the pumping efficiency is higher (Figure \ref{fig:figure3}(a)). However, circular polarization can also induce spin dynamics, with the precession amplitude reduced by only a factor of $\sqrt{2}$ compared to the linear case (Figure \ref{fig:figure3}(b)). The Fourier frequency spectrum of $\theta_1(t)$ and $\phi_1(t)$ in the case of short pulses is presented in Supplementary Section \ref{appendix_spectrum}.

For any polarization the amplitude of free precession excited by short pulses depends on the pulse duration $\tau$ non-monotonically, unlike the monotonic $\tau$-dependence for the case of long pulses (Fig. \ref{fig:figure4}). There is an optimal value of the pulse duration which gives maximum amplitude of the spin precession. The pumping of the magnetization oscillations is most efficient when two conditions are met: firstly, the pulse duration satisfies $\tau_{max} = \sqrt{2}/\omega$ (see the maximum of the curves in the Fig. \ref{fig:figure4}), and secondly, the magnetic field $\mathbf{h}$ is perpendicular to the external magnetic field, i.e. is along the $x$-axis, with phase $\varphi = 0$ (see Fig. \ref{fig:figure4}). 
The maximum amplitude of the precession angle $\phi_1$ for short pulses with $\tau_{max}$ equals to $\phi_{1max}=\frac{\gamma h_0\alpha}{\omega} \sqrt{\frac{2\pi\omega_1}{\omega_2}}\,\,e^{-1/2}$. Thus, for $\omega/2\pi=0.8$ THz optimal pulse duration is $\tau_{max} = 0.28$~ps and it provides oscillation amplitude $\phi_{1max}=6.5~$mrad for the linear polarization. Meanwhile similar amplitude can be also achieved by long pulses of circular polarization of the same amplitude if their duration is 40~ps.

\section{Conclusion}

In this work, we have developed a unified theoretical framework for describing the excitation of spin precession via Zeeman interaction in magnetic materials by high frequency pulses of arbitrary polarization with temporal Gaussian profile. Our analysis covers both the regimes of long (at least several oscillations within the pulse width) and short (less than one oscillation within the pulse width) pulses, taking into account the effects of the uniaxial magnetic anisotropy and the external magnetic field.

In the regime of long pulses ($\tau\omega \gg 2\pi$), a circularly polarized magnetic field acts as an effective rectified (non-oscillating) magnetic field along the $z$ axis. This leads to efficient excitation of spin precession via the Zeeman interaction, with the amplitude growing proportionally to $1/\omega$ and with increase of $\tau$. However, it is not possible to excite any long living free precession by linear polarized pulses.

In the regime of short pulses ($\tau\omega < 2\pi$) pulses of any polarization, including linear one can excite free spin precession. However, the dependence of the precession amplitude on the pulse duration $\tau$ becomes non-monotonic. There exists an optimal pulse duration at which the amplitude reaches its maximum for any polarization. The optimal pulse duration depends on magnetic parameters of the sample and the external magnetic field, as well as on the carrier frequency of the pulse and its amplitude. Therefore, it must be adjusted for each particular experiment, accordingly. This effect is particularly important for experiments with THz pulses, which often operate in this regime.

\section{Acknowledgments}
This work was financially supported by Russian Science Foundation (project № 23-62-10024). NIG and VIB also acknowledge support by BASIS foundation.

\bibliography{file}

\begin{thebibliography}{28}%
\makeatletter
\providecommand \@ifxundefined [1]{%
 \@ifx{#1\undefined}
}%
\providecommand \@ifnum [1]{%
 \ifnum #1\expandafter \@firstoftwo
 \else \expandafter \@secondoftwo
 \fi
}%
\providecommand \@ifx [1]{%
 \ifx #1\expandafter \@firstoftwo
 \else \expandafter \@secondoftwo
 \fi
}%
\providecommand \natexlab [1]{#1}%
\providecommand \enquote  [1]{``#1''}%
\providecommand \bibnamefont  [1]{#1}%
\providecommand \bibfnamefont [1]{#1}%
\providecommand \citenamefont [1]{#1}%
\providecommand \href@noop [0]{\@secondoftwo}%
\providecommand \href [0]{\begingroup \@sanitize@url \@href}%
\providecommand \@href[1]{\@@startlink{#1}\@@href}%
\providecommand \@@href[1]{\endgroup#1\@@endlink}%
\providecommand \@sanitize@url [0]{\catcode `\\12\catcode `\$12\catcode
  `\&12\catcode `\#12\catcode `\^12\catcode `\_12\catcode `\%12\relax}%
\providecommand \@@startlink[1]{}%
\providecommand \@@endlink[0]{}%
\providecommand \url  [0]{\begingroup\@sanitize@url \@url }%
\providecommand \@url [1]{\endgroup\@href {#1}{\urlprefix }}%
\providecommand \urlprefix  [0]{URL }%
\providecommand \Eprint [0]{\href }%
\providecommand \doibase [0]{https://doi.org/}%
\providecommand \selectlanguage [0]{\@gobble}%
\providecommand \bibinfo  [0]{\@secondoftwo}%
\providecommand \bibfield  [0]{\@secondoftwo}%
\providecommand \translation [1]{[#1]}%
\providecommand \BibitemOpen [0]{}%
\providecommand \bibitemStop [0]{}%
\providecommand \bibitemNoStop [0]{.\EOS\space}%
\providecommand \EOS [0]{\spacefactor3000\relax}%
\providecommand \BibitemShut  [1]{\csname bibitem#1\endcsname}%
\let\auto@bib@innerbib\@empty
\bibitem [{\citenamefont {Lu}\ \emph {et~al.}(2017)\citenamefont {Lu},
  \citenamefont {Li}, \citenamefont {Hwang}, \citenamefont {Ofori-Okai},
  \citenamefont {Kurihara}, \citenamefont {Suemoto},\ and\ \citenamefont
  {Nelson}}]{lu2017coherent}%
  \BibitemOpen
  \bibfield  {author} {\bibinfo {author} {\bibfnamefont {J.}~\bibnamefont
  {Lu}}, \bibinfo {author} {\bibfnamefont {X.}~\bibnamefont {Li}}, \bibinfo
  {author} {\bibfnamefont {H.~Y.}\ \bibnamefont {Hwang}}, \bibinfo {author}
  {\bibfnamefont {B.~K.}\ \bibnamefont {Ofori-Okai}}, \bibinfo {author}
  {\bibfnamefont {T.}~\bibnamefont {Kurihara}}, \bibinfo {author}
  {\bibfnamefont {T.}~\bibnamefont {Suemoto}},\ and\ \bibinfo {author}
  {\bibfnamefont {K.~A.}\ \bibnamefont {Nelson}},\ }\bibfield  {title}
  {\bibinfo {title} {Coherent two-dimensional terahertz magnetic resonance
  spectroscopy of collective spin waves},\ }\href@noop {} {\bibfield  {journal}
  {\bibinfo  {journal} {Physical review letters}\ }\textbf {\bibinfo {volume}
  {118}},\ \bibinfo {pages} {207204} (\bibinfo {year} {2017})}\BibitemShut
  {NoStop}%
\bibitem [{\citenamefont {Bocklage}(2016)}]{bocklage2016model}%
  \BibitemOpen
  \bibfield  {author} {\bibinfo {author} {\bibfnamefont {L.}~\bibnamefont
  {Bocklage}},\ }\bibfield  {title} {\bibinfo {title} {Model of thz
  magnetization dynamics},\ }\href@noop {} {\bibfield  {journal} {\bibinfo
  {journal} {Scientific reports}\ }\textbf {\bibinfo {volume} {6}},\ \bibinfo
  {pages} {22767} (\bibinfo {year} {2016})}\BibitemShut {NoStop}%
\bibitem [{\citenamefont {Mikhaylovskiy}\ \emph {et~al.}(2015)\citenamefont
  {Mikhaylovskiy}, \citenamefont {Huisman}, \citenamefont {Popov},
  \citenamefont {Zvezdin}, \citenamefont {Rasing}, \citenamefont {Pisarev},\
  and\ \citenamefont {Kimel}}]{mikhaylovskiy2015terahertz}%
  \BibitemOpen
  \bibfield  {author} {\bibinfo {author} {\bibfnamefont {R.}~\bibnamefont
  {Mikhaylovskiy}}, \bibinfo {author} {\bibfnamefont {T.}~\bibnamefont
  {Huisman}}, \bibinfo {author} {\bibfnamefont {A.}~\bibnamefont {Popov}},
  \bibinfo {author} {\bibfnamefont {A.}~\bibnamefont {Zvezdin}}, \bibinfo
  {author} {\bibfnamefont {T.}~\bibnamefont {Rasing}}, \bibinfo {author}
  {\bibfnamefont {R.}~\bibnamefont {Pisarev}},\ and\ \bibinfo {author}
  {\bibfnamefont {A.}~\bibnamefont {Kimel}},\ }\bibfield  {title} {\bibinfo
  {title} {Terahertz magnetization dynamics induced by femtosecond resonant
  pumping of dy 3+ subsystem in the multisublattice antiferromagnet dyfeo 3},\
  }\href@noop {} {\bibfield  {journal} {\bibinfo  {journal} {Physical Review
  B}\ }\textbf {\bibinfo {volume} {92}},\ \bibinfo {pages} {094437} (\bibinfo
  {year} {2015})}\BibitemShut {NoStop}%
\bibitem [{\citenamefont {Mikhaylovskiy}\ \emph {et~al.}(2016)\citenamefont
  {Mikhaylovskiy}, \citenamefont {Subkhangulov}, \citenamefont {Rasing},\ and\
  \citenamefont {Kimel}}]{mikhaylovskiy2016colossal}%
  \BibitemOpen
  \bibfield  {author} {\bibinfo {author} {\bibfnamefont {R.~V.}\ \bibnamefont
  {Mikhaylovskiy}}, \bibinfo {author} {\bibfnamefont {R.~R.}\ \bibnamefont
  {Subkhangulov}}, \bibinfo {author} {\bibfnamefont {T.}~\bibnamefont
  {Rasing}},\ and\ \bibinfo {author} {\bibfnamefont {A.~V.}\ \bibnamefont
  {Kimel}},\ }\bibfield  {title} {\bibinfo {title} {Colossal magneto-optical
  modulation at terahertz frequencies by counterpropagating femtosecond laser
  pulses in tb3ga5o12},\ }\href@noop {} {\bibfield  {journal} {\bibinfo
  {journal} {Optics letters}\ }\textbf {\bibinfo {volume} {41}},\ \bibinfo
  {pages} {5071} (\bibinfo {year} {2016})}\BibitemShut {NoStop}%
\bibitem [{\citenamefont {Blank}\ \emph {et~al.}(2021)\citenamefont {Blank},
  \citenamefont {Grishunin}, \citenamefont {Mashkovich}, \citenamefont
  {Logunov}, \citenamefont {Zvezdin},\ and\ \citenamefont
  {Kimel}}]{blank2021thz}%
  \BibitemOpen
  \bibfield  {author} {\bibinfo {author} {\bibfnamefont {T.~G.}\ \bibnamefont
  {Blank}}, \bibinfo {author} {\bibfnamefont {K.}~\bibnamefont {Grishunin}},
  \bibinfo {author} {\bibfnamefont {E.}~\bibnamefont {Mashkovich}}, \bibinfo
  {author} {\bibfnamefont {M.}~\bibnamefont {Logunov}}, \bibinfo {author}
  {\bibfnamefont {A.}~\bibnamefont {Zvezdin}},\ and\ \bibinfo {author}
  {\bibfnamefont {A.}~\bibnamefont {Kimel}},\ }\bibfield  {title} {\bibinfo
  {title} {Thz-scale field-induced spin dynamics in ferrimagnetic iron
  garnets},\ }\href@noop {} {\bibfield  {journal} {\bibinfo  {journal}
  {Physical Review Letters}\ }\textbf {\bibinfo {volume} {127}},\ \bibinfo
  {pages} {037203} (\bibinfo {year} {2021})}\BibitemShut {NoStop}%
\bibitem [{\citenamefont {Grishunin}\ \emph {et~al.}(2023)\citenamefont
  {Grishunin}, \citenamefont {Bilyk}, \citenamefont {Mishina}, \citenamefont
  {Kimel},\ and\ \citenamefont {Mashkovich}}]{grishunin2023two}%
  \BibitemOpen
  \bibfield  {author} {\bibinfo {author} {\bibfnamefont {K.}~\bibnamefont
  {Grishunin}}, \bibinfo {author} {\bibfnamefont {V.}~\bibnamefont {Bilyk}},
  \bibinfo {author} {\bibfnamefont {E.}~\bibnamefont {Mishina}}, \bibinfo
  {author} {\bibfnamefont {A.}~\bibnamefont {Kimel}},\ and\ \bibinfo {author}
  {\bibfnamefont {E.}~\bibnamefont {Mashkovich}},\ }\bibfield  {title}
  {\bibinfo {title} {Two-dimensional terahertz spectroscopy as a tool for
  revealing nonlinear interactions in media},\ }\href@noop {} {\bibfield
  {journal} {\bibinfo  {journal} {Review of Scientific Instruments}\ }\textbf
  {\bibinfo {volume} {94}} (\bibinfo {year} {2023})}\BibitemShut {NoStop}%
\bibitem [{\citenamefont {Shalaby}\ \emph {et~al.}(2016)\citenamefont
  {Shalaby}, \citenamefont {Vicario},\ and\ \citenamefont
  {Hauri}}]{shalaby2016simultaneous}%
  \BibitemOpen
  \bibfield  {author} {\bibinfo {author} {\bibfnamefont {M.}~\bibnamefont
  {Shalaby}}, \bibinfo {author} {\bibfnamefont {C.}~\bibnamefont {Vicario}},\
  and\ \bibinfo {author} {\bibfnamefont {C.~P.}\ \bibnamefont {Hauri}},\
  }\bibfield  {title} {\bibinfo {title} {Simultaneous electronic and the
  magnetic excitation of a ferromagnet by intense thz pulses},\ }\href@noop {}
  {\bibfield  {journal} {\bibinfo  {journal} {New Journal of Physics}\ }\textbf
  {\bibinfo {volume} {18}},\ \bibinfo {pages} {013019} (\bibinfo {year}
  {2016})}\BibitemShut {NoStop}%
\bibitem [{\citenamefont {Vicario}\ \emph {et~al.}(2013)\citenamefont
  {Vicario}, \citenamefont {Ruchert}, \citenamefont {Ardana-Lamas},
  \citenamefont {Derlet}, \citenamefont {Tudu}, \citenamefont {Luning},\ and\
  \citenamefont {Hauri}}]{vicario2013off}%
  \BibitemOpen
  \bibfield  {author} {\bibinfo {author} {\bibfnamefont {C.}~\bibnamefont
  {Vicario}}, \bibinfo {author} {\bibfnamefont {C.}~\bibnamefont {Ruchert}},
  \bibinfo {author} {\bibfnamefont {F.}~\bibnamefont {Ardana-Lamas}}, \bibinfo
  {author} {\bibfnamefont {P.~M.}\ \bibnamefont {Derlet}}, \bibinfo {author}
  {\bibfnamefont {B.}~\bibnamefont {Tudu}}, \bibinfo {author} {\bibfnamefont
  {J.}~\bibnamefont {Luning}},\ and\ \bibinfo {author} {\bibfnamefont {C.~P.}\
  \bibnamefont {Hauri}},\ }\bibfield  {title} {\bibinfo {title} {Off-resonant
  magnetization dynamics phase-locked to an intense phase-stable terahertz
  transient},\ }\href@noop {} {\bibfield  {journal} {\bibinfo  {journal}
  {Nature Photonics}\ }\textbf {\bibinfo {volume} {7}},\ \bibinfo {pages} {720}
  (\bibinfo {year} {2013})}\BibitemShut {NoStop}%
\bibitem [{\citenamefont {Unikandanunni}\ \emph {et~al.}(2022)\citenamefont
  {Unikandanunni}, \citenamefont {Medapalli}, \citenamefont {Asa},
  \citenamefont {Albisetti}, \citenamefont {Petti}, \citenamefont {Bertacco},
  \citenamefont {Fullerton},\ and\ \citenamefont
  {Bonetti}}]{unikandanunni2022inertial}%
  \BibitemOpen
  \bibfield  {author} {\bibinfo {author} {\bibfnamefont {V.}~\bibnamefont
  {Unikandanunni}}, \bibinfo {author} {\bibfnamefont {R.}~\bibnamefont
  {Medapalli}}, \bibinfo {author} {\bibfnamefont {M.}~\bibnamefont {Asa}},
  \bibinfo {author} {\bibfnamefont {E.}~\bibnamefont {Albisetti}}, \bibinfo
  {author} {\bibfnamefont {D.}~\bibnamefont {Petti}}, \bibinfo {author}
  {\bibfnamefont {R.}~\bibnamefont {Bertacco}}, \bibinfo {author}
  {\bibfnamefont {E.~E.}\ \bibnamefont {Fullerton}},\ and\ \bibinfo {author}
  {\bibfnamefont {S.}~\bibnamefont {Bonetti}},\ }\bibfield  {title} {\bibinfo
  {title} {Inertial spin dynamics in epitaxial cobalt films},\ }\href@noop {}
  {\bibfield  {journal} {\bibinfo  {journal} {Physical review letters}\
  }\textbf {\bibinfo {volume} {129}},\ \bibinfo {pages} {237201} (\bibinfo
  {year} {2022})}\BibitemShut {NoStop}%
\bibitem [{\citenamefont {Kampfrath}\ \emph {et~al.}(2011)\citenamefont
  {Kampfrath}, \citenamefont {Sell}, \citenamefont {Klatt}, \citenamefont
  {Pashkin}, \citenamefont {M{\"a}hrlein}, \citenamefont {Dekorsy},
  \citenamefont {Wolf}, \citenamefont {Fiebig}, \citenamefont {Leitenstorfer},\
  and\ \citenamefont {Huber}}]{kampfrath2011coherent}%
  \BibitemOpen
  \bibfield  {author} {\bibinfo {author} {\bibfnamefont {T.}~\bibnamefont
  {Kampfrath}}, \bibinfo {author} {\bibfnamefont {A.}~\bibnamefont {Sell}},
  \bibinfo {author} {\bibfnamefont {G.}~\bibnamefont {Klatt}}, \bibinfo
  {author} {\bibfnamefont {A.}~\bibnamefont {Pashkin}}, \bibinfo {author}
  {\bibfnamefont {S.}~\bibnamefont {M{\"a}hrlein}}, \bibinfo {author}
  {\bibfnamefont {T.}~\bibnamefont {Dekorsy}}, \bibinfo {author} {\bibfnamefont
  {M.}~\bibnamefont {Wolf}}, \bibinfo {author} {\bibfnamefont {M.}~\bibnamefont
  {Fiebig}}, \bibinfo {author} {\bibfnamefont {A.}~\bibnamefont
  {Leitenstorfer}},\ and\ \bibinfo {author} {\bibfnamefont {R.}~\bibnamefont
  {Huber}},\ }\bibfield  {title} {\bibinfo {title} {Coherent terahertz control
  of antiferromagnetic spin waves},\ }\href@noop {} {\bibfield  {journal}
  {\bibinfo  {journal} {Nature Photonics}\ }\textbf {\bibinfo {volume} {5}},\
  \bibinfo {pages} {31} (\bibinfo {year} {2011})}\BibitemShut {NoStop}%
\bibitem [{\citenamefont {Jin}\ \emph {et~al.}(2013)\citenamefont {Jin},
  \citenamefont {Mics}, \citenamefont {Ma}, \citenamefont {Cheng},
  \citenamefont {Bonn},\ and\ \citenamefont {Turchinovich}}]{jin2013single}%
  \BibitemOpen
  \bibfield  {author} {\bibinfo {author} {\bibfnamefont {Z.}~\bibnamefont
  {Jin}}, \bibinfo {author} {\bibfnamefont {Z.}~\bibnamefont {Mics}}, \bibinfo
  {author} {\bibfnamefont {G.}~\bibnamefont {Ma}}, \bibinfo {author}
  {\bibfnamefont {Z.}~\bibnamefont {Cheng}}, \bibinfo {author} {\bibfnamefont
  {M.}~\bibnamefont {Bonn}},\ and\ \bibinfo {author} {\bibfnamefont
  {D.}~\bibnamefont {Turchinovich}},\ }\bibfield  {title} {\bibinfo {title}
  {Single-pulse terahertz coherent control of spin resonance in the canted
  antiferromagnet yfeo 3, mediated by dielectric anisotropy},\ }\href@noop {}
  {\bibfield  {journal} {\bibinfo  {journal} {Physical Review B—Condensed
  Matter and Materials Physics}\ }\textbf {\bibinfo {volume} {87}},\ \bibinfo
  {pages} {094422} (\bibinfo {year} {2013})}\BibitemShut {NoStop}%
\bibitem [{\citenamefont {Baierl}\ \emph {et~al.}(2016)\citenamefont {Baierl},
  \citenamefont {Mentink}, \citenamefont {Hohenleutner}, \citenamefont {Braun},
  \citenamefont {Do}, \citenamefont {Lange}, \citenamefont {Sell},
  \citenamefont {Fiebig}, \citenamefont {Woltersdorf}, \citenamefont
  {Kampfrath} \emph {et~al.}}]{baierl2016terahertz}%
  \BibitemOpen
  \bibfield  {author} {\bibinfo {author} {\bibfnamefont {S.}~\bibnamefont
  {Baierl}}, \bibinfo {author} {\bibfnamefont {J.~H.}\ \bibnamefont {Mentink}},
  \bibinfo {author} {\bibfnamefont {M.}~\bibnamefont {Hohenleutner}}, \bibinfo
  {author} {\bibfnamefont {L.}~\bibnamefont {Braun}}, \bibinfo {author}
  {\bibfnamefont {T.-M.}\ \bibnamefont {Do}}, \bibinfo {author} {\bibfnamefont
  {C.}~\bibnamefont {Lange}}, \bibinfo {author} {\bibfnamefont
  {A.}~\bibnamefont {Sell}}, \bibinfo {author} {\bibfnamefont {M.}~\bibnamefont
  {Fiebig}}, \bibinfo {author} {\bibfnamefont {G.}~\bibnamefont {Woltersdorf}},
  \bibinfo {author} {\bibfnamefont {T.}~\bibnamefont {Kampfrath}}, \emph
  {et~al.},\ }\bibfield  {title} {\bibinfo {title} {Terahertz-driven nonlinear
  spin response of antiferromagnetic nickel oxide},\ }\href@noop {} {\bibfield
  {journal} {\bibinfo  {journal} {Physical review letters}\ }\textbf {\bibinfo
  {volume} {117}},\ \bibinfo {pages} {197201} (\bibinfo {year}
  {2016})}\BibitemShut {NoStop}%
\bibitem [{\citenamefont {Mashkovich}\ \emph {et~al.}(2019)\citenamefont
  {Mashkovich}, \citenamefont {Grishunin}, \citenamefont {Mikhaylovskiy},
  \citenamefont {Zvezdin}, \citenamefont {Pisarev}, \citenamefont {Strugatsky},
  \citenamefont {Christianen}, \citenamefont {Rasing},\ and\ \citenamefont
  {Kimel}}]{mashkovich2019terahertz}%
  \BibitemOpen
  \bibfield  {author} {\bibinfo {author} {\bibfnamefont {E.}~\bibnamefont
  {Mashkovich}}, \bibinfo {author} {\bibfnamefont {K.}~\bibnamefont
  {Grishunin}}, \bibinfo {author} {\bibfnamefont {R.}~\bibnamefont
  {Mikhaylovskiy}}, \bibinfo {author} {\bibfnamefont {A.}~\bibnamefont
  {Zvezdin}}, \bibinfo {author} {\bibfnamefont {R.}~\bibnamefont {Pisarev}},
  \bibinfo {author} {\bibfnamefont {M.}~\bibnamefont {Strugatsky}}, \bibinfo
  {author} {\bibfnamefont {P.}~\bibnamefont {Christianen}}, \bibinfo {author}
  {\bibfnamefont {T.}~\bibnamefont {Rasing}},\ and\ \bibinfo {author}
  {\bibfnamefont {A.}~\bibnamefont {Kimel}},\ }\bibfield  {title} {\bibinfo
  {title} {Terahertz optomagnetism: nonlinear thz excitation of ghz spin waves
  in antiferromagnetic febo 3},\ }\href@noop {} {\bibfield  {journal} {\bibinfo
   {journal} {Physical review letters}\ }\textbf {\bibinfo {volume} {123}},\
  \bibinfo {pages} {157202} (\bibinfo {year} {2019})}\BibitemShut {NoStop}%
\bibitem [{\citenamefont {Grishunin}\ \emph {et~al.}(2021)\citenamefont
  {Grishunin}, \citenamefont {Mashkovich}, \citenamefont {Kimel}, \citenamefont
  {Balbashov},\ and\ \citenamefont {Zvezdin}}]{grishunin2021excitation}%
  \BibitemOpen
  \bibfield  {author} {\bibinfo {author} {\bibfnamefont {K.}~\bibnamefont
  {Grishunin}}, \bibinfo {author} {\bibfnamefont {E.}~\bibnamefont
  {Mashkovich}}, \bibinfo {author} {\bibfnamefont {A.}~\bibnamefont {Kimel}},
  \bibinfo {author} {\bibfnamefont {A.}~\bibnamefont {Balbashov}},\ and\
  \bibinfo {author} {\bibfnamefont {A.}~\bibnamefont {Zvezdin}},\ }\bibfield
  {title} {\bibinfo {title} {Excitation and detection of terahertz coherent
  spin waves in antiferromagnetic $\alpha$-fe 2 o 3},\ }\href@noop {}
  {\bibfield  {journal} {\bibinfo  {journal} {Physical Review B}\ }\textbf
  {\bibinfo {volume} {104}},\ \bibinfo {pages} {024419} (\bibinfo {year}
  {2021})}\BibitemShut {NoStop}%
\bibitem [{\citenamefont {Blank}\ \emph {et~al.}(2023)\citenamefont {Blank},
  \citenamefont {Grishunin}, \citenamefont {Ivanov}, \citenamefont
  {Mashkovich}, \citenamefont {Afanasiev},\ and\ \citenamefont
  {Kimel}}]{blank2023empowering}%
  \BibitemOpen
  \bibfield  {author} {\bibinfo {author} {\bibfnamefont {T.~G.}\ \bibnamefont
  {Blank}}, \bibinfo {author} {\bibfnamefont {K.~A.}\ \bibnamefont
  {Grishunin}}, \bibinfo {author} {\bibfnamefont {B.~A.}\ \bibnamefont
  {Ivanov}}, \bibinfo {author} {\bibfnamefont {E.~A.}\ \bibnamefont
  {Mashkovich}}, \bibinfo {author} {\bibfnamefont {D.}~\bibnamefont
  {Afanasiev}},\ and\ \bibinfo {author} {\bibfnamefont {A.~V.}\ \bibnamefont
  {Kimel}},\ }\bibfield  {title} {\bibinfo {title} {Empowering control of
  antiferromagnets by thz-induced spin coherence},\ }\href@noop {} {\bibfield
  {journal} {\bibinfo  {journal} {Physical Review Letters}\ }\textbf {\bibinfo
  {volume} {131}},\ \bibinfo {pages} {096701} (\bibinfo {year}
  {2023})}\BibitemShut {NoStop}%
\bibitem [{\citenamefont {Kampfrath}\ \emph {et~al.}(2013)\citenamefont
  {Kampfrath}, \citenamefont {Tanaka},\ and\ \citenamefont
  {Nelson}}]{kampfrath2013resonant}%
  \BibitemOpen
  \bibfield  {author} {\bibinfo {author} {\bibfnamefont {T.}~\bibnamefont
  {Kampfrath}}, \bibinfo {author} {\bibfnamefont {K.}~\bibnamefont {Tanaka}},\
  and\ \bibinfo {author} {\bibfnamefont {K.~A.}\ \bibnamefont {Nelson}},\
  }\bibfield  {title} {\bibinfo {title} {Resonant and nonresonant control over
  matter and light by intense terahertz transients},\ }\href@noop {} {\bibfield
   {journal} {\bibinfo  {journal} {Nature Photonics}\ }\textbf {\bibinfo
  {volume} {7}},\ \bibinfo {pages} {680} (\bibinfo {year} {2013})}\BibitemShut
  {NoStop}%
\bibitem [{\citenamefont {Vicario}\ \emph {et~al.}(2014)\citenamefont
  {Vicario}, \citenamefont {Ruchert}, \citenamefont {Ardana-Lamas},
  \citenamefont {Derlet}, \citenamefont {Tudu}, \citenamefont {Luning},\ and\
  \citenamefont {Hauri}}]{vicario2014terahertz}%
  \BibitemOpen
  \bibfield  {author} {\bibinfo {author} {\bibfnamefont {C.}~\bibnamefont
  {Vicario}}, \bibinfo {author} {\bibfnamefont {C.}~\bibnamefont {Ruchert}},
  \bibinfo {author} {\bibfnamefont {F.}~\bibnamefont {Ardana-Lamas}}, \bibinfo
  {author} {\bibfnamefont {P.}~\bibnamefont {Derlet}}, \bibinfo {author}
  {\bibfnamefont {B.}~\bibnamefont {Tudu}}, \bibinfo {author} {\bibfnamefont
  {J.}~\bibnamefont {Luning}},\ and\ \bibinfo {author} {\bibfnamefont
  {C.}~\bibnamefont {Hauri}},\ }\bibfield  {title} {\bibinfo {title}
  {Terahertz-driven non-resonant magnetization dynamics in cobalt},\ }in\
  \href@noop {} {\emph {\bibinfo {booktitle} {CLEO: Science and Innovations}}}\
  (\bibinfo {organization} {Optica Publishing Group},\ \bibinfo {year} {2014})\
  pp.\ \bibinfo {pages} {SF2J--1}\BibitemShut {NoStop}%
\bibitem [{\citenamefont {Rongione}\ \emph {et~al.}(2023)\citenamefont
  {Rongione}, \citenamefont {Gueckstock}, \citenamefont {Mattern},
  \citenamefont {Gomonay}, \citenamefont {Meer}, \citenamefont {Schmitt},
  \citenamefont {Ramos}, \citenamefont {Kikkawa}, \citenamefont {Mi{\v{c}}ica},
  \citenamefont {Saitoh} \emph {et~al.}}]{rongione2023emission}%
  \BibitemOpen
  \bibfield  {author} {\bibinfo {author} {\bibfnamefont {E.}~\bibnamefont
  {Rongione}}, \bibinfo {author} {\bibfnamefont {O.}~\bibnamefont
  {Gueckstock}}, \bibinfo {author} {\bibfnamefont {M.}~\bibnamefont {Mattern}},
  \bibinfo {author} {\bibfnamefont {O.}~\bibnamefont {Gomonay}}, \bibinfo
  {author} {\bibfnamefont {H.}~\bibnamefont {Meer}}, \bibinfo {author}
  {\bibfnamefont {C.}~\bibnamefont {Schmitt}}, \bibinfo {author} {\bibfnamefont
  {R.}~\bibnamefont {Ramos}}, \bibinfo {author} {\bibfnamefont
  {T.}~\bibnamefont {Kikkawa}}, \bibinfo {author} {\bibfnamefont
  {M.}~\bibnamefont {Mi{\v{c}}ica}}, \bibinfo {author} {\bibfnamefont
  {E.}~\bibnamefont {Saitoh}}, \emph {et~al.},\ }\bibfield  {title} {\bibinfo
  {title} {Emission of coherent thz magnons in an antiferromagnetic insulator
  triggered by ultrafast spin--phonon interactions},\ }\href@noop {} {\bibfield
   {journal} {\bibinfo  {journal} {Nature communications}\ }\textbf {\bibinfo
  {volume} {14}},\ \bibinfo {pages} {1818} (\bibinfo {year}
  {2023})}\BibitemShut {NoStop}%
\bibitem [{\citenamefont {Hudl}\ \emph {et~al.}(2019)\citenamefont {Hudl},
  \citenamefont {d’Aquino}, \citenamefont {Pancaldi}, \citenamefont {Yang},
  \citenamefont {Samant}, \citenamefont {Parkin}, \citenamefont {D{\"u}rr},
  \citenamefont {Serpico}, \citenamefont {Hoffmann},\ and\ \citenamefont
  {Bonetti}}]{hudl2019nonlinear}%
  \BibitemOpen
  \bibfield  {author} {\bibinfo {author} {\bibfnamefont {M.}~\bibnamefont
  {Hudl}}, \bibinfo {author} {\bibfnamefont {M.}~\bibnamefont {d’Aquino}},
  \bibinfo {author} {\bibfnamefont {M.}~\bibnamefont {Pancaldi}}, \bibinfo
  {author} {\bibfnamefont {S.-H.}\ \bibnamefont {Yang}}, \bibinfo {author}
  {\bibfnamefont {M.~G.}\ \bibnamefont {Samant}}, \bibinfo {author}
  {\bibfnamefont {S.~S.}\ \bibnamefont {Parkin}}, \bibinfo {author}
  {\bibfnamefont {H.~A.}\ \bibnamefont {D{\"u}rr}}, \bibinfo {author}
  {\bibfnamefont {C.}~\bibnamefont {Serpico}}, \bibinfo {author} {\bibfnamefont
  {M.~C.}\ \bibnamefont {Hoffmann}},\ and\ \bibinfo {author} {\bibfnamefont
  {S.}~\bibnamefont {Bonetti}},\ }\bibfield  {title} {\bibinfo {title}
  {Nonlinear magnetization dynamics driven by strong terahertz fields},\
  }\href@noop {} {\bibfield  {journal} {\bibinfo  {journal} {Physical review
  letters}\ }\textbf {\bibinfo {volume} {123}},\ \bibinfo {pages} {197204}
  (\bibinfo {year} {2019})}\BibitemShut {NoStop}%
\bibitem [{\citenamefont {Shalaby}\ \emph {et~al.}(2017)\citenamefont
  {Shalaby}, \citenamefont {Vicario}, \citenamefont {Giorgianni}, \citenamefont
  {Donges}, \citenamefont {Carva}, \citenamefont {Oppeneer}, \citenamefont
  {Nowak},\ and\ \citenamefont {Hauri}}]{shalaby2017off}%
  \BibitemOpen
  \bibfield  {author} {\bibinfo {author} {\bibfnamefont {M.}~\bibnamefont
  {Shalaby}}, \bibinfo {author} {\bibfnamefont {C.}~\bibnamefont {Vicario}},
  \bibinfo {author} {\bibfnamefont {F.}~\bibnamefont {Giorgianni}}, \bibinfo
  {author} {\bibfnamefont {A.}~\bibnamefont {Donges}}, \bibinfo {author}
  {\bibfnamefont {K.}~\bibnamefont {Carva}}, \bibinfo {author} {\bibfnamefont
  {P.~M.}\ \bibnamefont {Oppeneer}}, \bibinfo {author} {\bibfnamefont
  {U.}~\bibnamefont {Nowak}},\ and\ \bibinfo {author} {\bibfnamefont {C.~P.}\
  \bibnamefont {Hauri}},\ }\bibfield  {title} {\bibinfo {title} {Off-resonant
  magnetization dynamics in co, fe and ni thin films driven by an intense
  single-cycle thz field},\ }in\ \href@noop {} {\emph {\bibinfo {booktitle}
  {CLEO: Science and Innovations}}}\ (\bibinfo {organization} {Optica
  Publishing Group},\ \bibinfo {year} {2017})\ pp.\ \bibinfo {pages}
  {STu1J--1}\BibitemShut {NoStop}%
\bibitem [{\citenamefont {F{\"u}l{\"o}p}\ \emph {et~al.}(2020)\citenamefont
  {F{\"u}l{\"o}p}, \citenamefont {Tzortzakis},\ and\ \citenamefont
  {Kampfrath}}]{fulop2020laser}%
  \BibitemOpen
  \bibfield  {author} {\bibinfo {author} {\bibfnamefont {J.~A.}\ \bibnamefont
  {F{\"u}l{\"o}p}}, \bibinfo {author} {\bibfnamefont {S.}~\bibnamefont
  {Tzortzakis}},\ and\ \bibinfo {author} {\bibfnamefont {T.}~\bibnamefont
  {Kampfrath}},\ }\bibfield  {title} {\bibinfo {title} {Laser-driven
  strong-field terahertz sources},\ }\href@noop {} {\bibfield  {journal}
  {\bibinfo  {journal} {Advanced Optical Materials}\ }\textbf {\bibinfo
  {volume} {8}},\ \bibinfo {pages} {1900681} (\bibinfo {year}
  {2020})}\BibitemShut {NoStop}%
\bibitem [{\citenamefont {Wu}\ \emph {et~al.}(2018)\citenamefont {Wu},
  \citenamefont {Ma}, \citenamefont {Zhang}, \citenamefont {Chai},
  \citenamefont {Fang}, \citenamefont {Xia}, \citenamefont {Kong},
  \citenamefont {Wang}, \citenamefont {Liu}, \citenamefont {Zhu} \emph
  {et~al.}}]{wu2018highly}%
  \BibitemOpen
  \bibfield  {author} {\bibinfo {author} {\bibfnamefont {X.-j.}\ \bibnamefont
  {Wu}}, \bibinfo {author} {\bibfnamefont {J.-l.}\ \bibnamefont {Ma}}, \bibinfo
  {author} {\bibfnamefont {B.-l.}\ \bibnamefont {Zhang}}, \bibinfo {author}
  {\bibfnamefont {S.-s.}\ \bibnamefont {Chai}}, \bibinfo {author}
  {\bibfnamefont {Z.-j.}\ \bibnamefont {Fang}}, \bibinfo {author}
  {\bibfnamefont {C.-Y.}\ \bibnamefont {Xia}}, \bibinfo {author} {\bibfnamefont
  {D.-y.}\ \bibnamefont {Kong}}, \bibinfo {author} {\bibfnamefont {J.-g.}\
  \bibnamefont {Wang}}, \bibinfo {author} {\bibfnamefont {H.}~\bibnamefont
  {Liu}}, \bibinfo {author} {\bibfnamefont {C.-Q.}\ \bibnamefont {Zhu}}, \emph
  {et~al.},\ }\bibfield  {title} {\bibinfo {title} {Highly efficient generation
  of 0.2 mj terahertz pulses in lithium niobate at room temperature with sub-50
  fs chirped ti: sapphire laser pulses},\ }\href@noop {} {\bibfield  {journal}
  {\bibinfo  {journal} {Optics express}\ }\textbf {\bibinfo {volume} {26}},\
  \bibinfo {pages} {7107} (\bibinfo {year} {2018})}\BibitemShut {NoStop}%
\bibitem [{\citenamefont {Krinchik}\ and\ \citenamefont
  {Chetkin}(1969)}]{krinchik1969transparent}%
  \BibitemOpen
  \bibfield  {author} {\bibinfo {author} {\bibfnamefont {G.~S.}\ \bibnamefont
  {Krinchik}}\ and\ \bibinfo {author} {\bibfnamefont {M.~V.}\ \bibnamefont
  {Chetkin}},\ }\bibfield  {title} {\bibinfo {title} {Transparent
  ferromagnets},\ }\href@noop {} {\bibfield  {journal} {\bibinfo  {journal}
  {Soviet physics uspekhi}\ }\textbf {\bibinfo {volume} {12}},\ \bibinfo
  {pages} {307} (\bibinfo {year} {1969})}\BibitemShut {NoStop}%
\bibitem [{\citenamefont {Gribova}\ \emph {et~al.}(2024)\citenamefont
  {Gribova}, \citenamefont {Berzhansky}, \citenamefont {Polulyakh},\ and\
  \citenamefont {Belotelov}}]{gribova2024inverse}%
  \BibitemOpen
  \bibfield  {author} {\bibinfo {author} {\bibfnamefont {N.}~\bibnamefont
  {Gribova}}, \bibinfo {author} {\bibfnamefont {V.}~\bibnamefont {Berzhansky}},
  \bibinfo {author} {\bibfnamefont {S.}~\bibnamefont {Polulyakh}},\ and\
  \bibinfo {author} {\bibfnamefont {V.}~\bibnamefont {Belotelov}},\ }\bibfield
  {title} {\bibinfo {title} {Inverse faraday effect in ferrite--garnet films in
  the near-infrared range},\ }\href@noop {} {\bibfield  {journal} {\bibinfo
  {journal} {JETP Letters}\ }\textbf {\bibinfo {volume} {120}},\ \bibinfo
  {pages} {183} (\bibinfo {year} {2024})}\BibitemShut {NoStop}%
\bibitem [{\citenamefont {Eleonskii}\ \emph {et~al.}(1977)\citenamefont
  {Eleonskii}, \citenamefont {Zvezdin},\ and\ \citenamefont
  {Redko}}]{eleonskii1977effect}%
  \BibitemOpen
  \bibfield  {author} {\bibinfo {author} {\bibfnamefont {V.}~\bibnamefont
  {Eleonskii}}, \bibinfo {author} {\bibfnamefont {A.}~\bibnamefont {Zvezdin}},\
  and\ \bibinfo {author} {\bibfnamefont {V.}~\bibnamefont {Redko}},\ }\bibfield
   {title} {\bibinfo {title} {Effect of rapidly oscillating magnetic fields on
  the domain structure of ferromagnetics},\ }\href@noop {} {\bibfield
  {journal} {\bibinfo  {journal} {Fiz. Met. Metalloved.}\ }\textbf {\bibinfo
  {volume} {43}},\ \bibinfo {pages} {7} (\bibinfo {year} {1977})}\BibitemShut
  {NoStop}%
\bibitem [{\citenamefont {Pitaevskii}(1961)}]{pitaevskii1961electric}%
  \BibitemOpen
  \bibfield  {author} {\bibinfo {author} {\bibfnamefont {L.}~\bibnamefont
  {Pitaevskii}},\ }\bibfield  {title} {\bibinfo {title} {Electric forces in a
  transparent dispersive medium},\ }\href@noop {} {\bibfield  {journal}
  {\bibinfo  {journal} {Sov. Phys. JETP}\ }\textbf {\bibinfo {volume} {12}},\
  \bibinfo {pages} {1008} (\bibinfo {year} {1961})}\BibitemShut {NoStop}%
\bibitem [{\citenamefont {Van~der Ziel}\ \emph {et~al.}(1965)\citenamefont
  {Van~der Ziel}, \citenamefont {Pershan},\ and\ \citenamefont
  {Malmstrom}}]{van1965optically}%
  \BibitemOpen
  \bibfield  {author} {\bibinfo {author} {\bibfnamefont {J.}~\bibnamefont
  {Van~der Ziel}}, \bibinfo {author} {\bibfnamefont {P.~S.}\ \bibnamefont
  {Pershan}},\ and\ \bibinfo {author} {\bibfnamefont {L.}~\bibnamefont
  {Malmstrom}},\ }\bibfield  {title} {\bibinfo {title} {Optically-induced
  magnetization resulting from the inverse faraday effect},\ }\href@noop {}
  {\bibfield  {journal} {\bibinfo  {journal} {Physical review letters}\
  }\textbf {\bibinfo {volume} {15}},\ \bibinfo {pages} {190} (\bibinfo {year}
  {1965})}\BibitemShut {NoStop}%
\bibitem [{\citenamefont {Pershan}\ \emph {et~al.}(1966)\citenamefont
  {Pershan}, \citenamefont {Van~der Ziel},\ and\ \citenamefont
  {Malmstrom}}]{pershan1966theoretical}%
  \BibitemOpen
  \bibfield  {author} {\bibinfo {author} {\bibfnamefont {P.}~\bibnamefont
  {Pershan}}, \bibinfo {author} {\bibfnamefont {J.}~\bibnamefont {Van~der
  Ziel}},\ and\ \bibinfo {author} {\bibfnamefont {L.}~\bibnamefont
  {Malmstrom}},\ }\bibfield  {title} {\bibinfo {title} {Theoretical discussion
  of the inverse faraday effect, raman scattering, and related phenomena},\
  }\href@noop {} {\bibfield  {journal} {\bibinfo  {journal} {Physical review}\
  }\textbf {\bibinfo {volume} {143}},\ \bibinfo {pages} {574} (\bibinfo {year}
  {1966})}\BibitemShut {NoStop}%
\end{thebibliography}%

\clearpage

\begin{widetext}
\section*{Supplementary}

\setcounter{section}{0}
\setcounter{equation}{0}
\setcounter{figure}{0}
\numberwithin{equation}{section}
\renewcommand\thesection{\Alph{section}}
\renewcommand\theequation{\Alph{section}.\arabic{equation}}
\renewcommand\thefigure{S.\arabic{figure}}
\section{Solution of the equations of motion in the limit of long pulses}\label{appendix1}
If we consider the effect of circularly polarized light on magnetization in a magneto-optical material, we can derive a general equation of motion for magnetization. These equations describe the magnetodipole contribution to the inverse Faraday effect. High frequency electro-magnetic field of light with arbitrary polarization $\mathbf{h} = \frac{1}{2}(\mathbf{h}_0 e^{i\omega t} +\mathbf{h}_0^{*} e^{-i\omega t})f(t)= \bigl(\text{Re}\mathbf{h}_0\cos\omega t - \text{Im}\mathbf{h}_0\sin\omega t\bigr)f(t) $, where $\omega = \frac{2\pi c}{\lambda}$ is frequency, $\tau$ is characteristic duration of pulse (the characteristic time at which the function is not zero), the envelope function $f(t)$ is slow on the scale $2\pi/\omega$. 

However, in order to obtain the time dependence of $H_{\mathrm{IFE}}(t)$, it is necessary to use the Lagrangian formalism, which assumes a description of magnetodynamics. Lagrangian in spherical coordinates of thin monodomain magnetooptical film with biaxial anisotropy is equal to:
\begin{equation}
\mathcal{L} = -\frac{M}{\gamma}\dot{\phi}\cos{\theta} - U_{1,2}  = 
-\frac{M}{\gamma}\dot{\phi}\cos{\theta} - K_1\cos^2\theta -K_2\sin^2\theta\cos^2\phi-V_h
\label{supp:lagrangian}
\end{equation}
where the first contribution to the right part in the middle equation is kinetic, the second is potential; $K_1$ and $K_2$ are the magnetic anisotropy constants, $V_h = -M(h_x\sin\theta\cos\phi+h_y\sin\theta\sin\phi)$ is Zeeman energy of the high frequency oscillations of electro-magnetic field. Let us derive the Euler-Lagrange equations for the Lagrangian (Eq. \ref{supp:lagrangian}) of a magnetic material with biaxial anisotropy  for generalized coordinates $(\theta, \phi, r)$:
\begin{align}
&\dot{\phi}\sin\theta+\cos\theta\sin\theta(\omega_1-\omega_2\cos^2\phi)-\frac{\gamma}{M}\frac{\partial V_h}{\partial \theta}=0,\label{supp:phi_dot}\\
&\dot{\theta}\sin\theta-\omega_2\sin^2\theta\cos\phi\sin\phi+\frac{\gamma}{M}\frac{\partial V_h}{\partial \phi} =0\label{supp:theta_dot}
\end{align}
where $\omega_1 = \frac{2K_1\gamma}{M}$ and $\omega_2 = \frac{2K_2\gamma}{M}$ in the case of biaxial anisotropy without constant external field; $\omega_1 = \frac{2K_1\gamma}{M}+\gamma H_{ext}$ and $\omega_2 = \gamma H_{ext}$ in the case of uniaxial anisotropy with external magnetic field along $y$ axis.

Ground state is the static state defined by relations $\partial U/\partial \theta = 0$ and $\partial U/\partial \phi = 0$. Assuming that $K_1, K_2 > 0$, the ground state of $\mathbf{M}$ is determined by $\phi_0 = \pi/2, 3\pi/2$ and $\theta_0 = \pi/2$. In our work $\phi = \frac{\pi}{2}+\phi_1$ and $\theta = \frac{\pi}{2}+\theta_1$ is selected as ground state with respect to $\theta_1, \phi_1 \ll 1$. In this paper, we work in the configuration shown in the Figure \ref{fig:scheme1}, and the electromagnetic field of light illuminate along the normal, therefore, since the waves are transverse, then $h_z = 0$. To analyze the dependence of the inverse magneto-optical effects on polarization, it is necessary to normalize electromagnetic field by energy, therefore, $h_x = h_0 (\alpha\cos\psi\cos\omega t+\sqrt{1-\alpha^2}\sin\psi\sin\omega t)f(t)$ and $h_y = h_0(-\alpha\sin\psi\cos\omega t+\sqrt{1-\alpha^2}\cos\psi\sin\omega t)f(t)$. Thus, the polarization of light is determined by two parameters $\alpha$ and $\psi$.

Simplifying the equations (\ref{supp:phi_dot}),~(\ref{supp:theta_dot}) and taking into account terms up to the first order in $\theta_1, \phi_1$ we have
\begin{align}
\dot{\phi}_1-\omega_1 \theta_1 &= \gamma h_y \theta_1,\label{supp:phi1_dot}\\
\dot{\theta}_1 + \omega_2 \phi_1&=-\gamma(h_x+h_y\phi_1).\label{supp:theta1_dot}
\end{align}

Under this condition $\omega_1, \omega_2 \ll \omega$, we divide equations \ref{supp:phi1_dot} and \ref{supp:theta1_dot} into smooth $\bar{\theta}_1$, $\bar{\phi}_1$ and rapidly $\theta_1'$, $\phi_1'$ oscillating parts. This method is correct when considering a limit $\frac{2\pi}{\omega}\ll\tau$, which we assume till the end of this section. Thus, with respect to $\phi_1 = \bar{\phi}_1 + \phi_1'$, $\theta_1 = \bar{\theta}_1 + \theta_1'$ we get

\begin{equation}
	\left\{\begin{aligned}
		&\dot{\phi}_1' - \omega_1\theta_1'= \gamma h_y \bar{\theta}_1\\
		&\dot{\theta}_1' + \omega_2\phi_1'=-\gamma(h_x+h_y\bar{\phi}_1)
	\end{aligned}\right.
    \qquad \qquad \qquad
        \left\{\begin{aligned}
		&\dot{\bar{\phi}}_1-\omega_1 {\bar{\theta}}_1 = \gamma h_y\theta_1'\\
		&\dot{\bar{\theta}}_1+\omega_2 {\bar{\phi}}_1=-\gamma h_y \phi_1'
	\end{aligned}\right.
    \label{eq:devision}
\end{equation}

Using the Green's function for equations with rapidly oscillating parts in Equations \ref{eq:devision} and after applying the integration by parts and neglecting the terms with smooth functions $\bar{\theta}_1(t)\ll1$ and $\bar{\phi}_1(t)\ll1$, in the case $\omega_1, \omega_2 \ll \omega$ we have:
\begin{align}
    &\phi_1'(t)  
    \approx
    -\frac{\gamma}{\omega}\sqrt{\frac{\omega_1}{\omega_2}}\sin(\omega_rt)\Bigr[\alpha\cos\psi\sin\omega t-\sqrt{1-\alpha^2}\sin\psi\cos\omega t\Bigl]f(t),\label{simp_rapid_phi}\\
    &\theta_1'(t) 
    \approx
     -\frac{\gamma}{\omega}\cos(\omega_rt)\Bigr[\alpha\cos\psi\sin\omega t-\sqrt{1-\alpha^2}\sin\psi\cos\omega t\Bigl]f(t),\label{simp_rapid_theta}
\end{align}
where $\omega_r = \sqrt{\omega_1\omega_2}$ is the resonant ferromagnetic frequency, the right part of the equations obtained in the limit $\frac{2\pi}{\omega}\ll\tau$.

Substituting rapidly oscillating functions (Eq. \ref{simp_rapid_phi} and \ref{simp_rapid_theta}) into the smooth equations (\ref{eq:devision}), we get the solution in following form
\vspace{-0.3cm}
\begin{equation}
\begin{aligned}
    \bar{\phi}_1(t) = -\frac{\gamma^2 \alpha \sqrt{1-\alpha^2}}{4\omega} \cos\omega_rt\Biggl[\Bigl(1+\frac{\omega_1}{\omega_2}\Bigr)&\!\!\!\int\limits_{-\infty}^{t}dt^\prime f^2(t^{\prime})+\Bigl(1-\frac{\omega_1}{\omega_2}\Bigr)\!\!\!\int\limits_{-\infty}^{t}dt^\prime f^2(t^{\prime})\cos2\omega_rt^{\prime}\Biggr]\\ &-
    \frac{\gamma^2 \alpha \sqrt{1-\alpha^2}}{4\omega} \sin\omega_rt \Bigl(1-\frac{\omega_1}{\omega_2}\Bigr)\!\!\!\int\limits_{-\infty}^{t}dt^\prime f^2(t^{\prime})\sin 2\omega_rt^{\prime}
   ,\label{phi_faraday}
\end{aligned}
\end{equation}
\vspace{-0.5cm}
\begin{equation}
\begin{aligned}
    \bar{\theta}_1(t) = \frac{\gamma^2 \alpha \sqrt{1-\alpha^2}}{4\omega} \sqrt{\frac{\omega_1}{\omega_2}}
    \sin\omega_rt
    \Biggl[&\Bigl(1+\frac{\omega_2}{\omega_1}\Bigr)\!\!\!\int\limits_{-\infty}^{t}dt^\prime f^2(t^{\prime})-\Bigl(1-\frac{\omega_2}{\omega_1}\Bigr)\!\!\!\int\limits_{-\infty}^{t}dt^\prime f^2(t^{\prime})\cos2\omega_rt^{\prime}\Biggr]\\ &+
    \frac{\gamma^2 \alpha \sqrt{1-\alpha^2}}{4\omega} \sqrt{\frac{\omega_1}{\omega_2}}\cos\omega_rt \Bigl(1-\frac{\omega_2}{\omega_1}\Bigr)\!\!\!\int\limits_{-\infty}^{t}dt^\prime f^2(t^{\prime})\sin 2\omega_rt^{\prime}
    ,\label{theta_faraday}
\end{aligned}
\end{equation}
where the dominant contribution comes from the integral of the square of the envelope function. The second and third terms in right part of the Equations \ref{phi_faraday} and \ref{theta_faraday} are suppressed since $\tau\gg2\pi/\omega$. 

In the case of gaussian profile $f(t) = e^{-\frac{t^2}{\tau^2}}$ and $\tau\gg 2\pi/\omega$ we can get a solution 
\begin{equation}
\begin{aligned}
    \bar{\phi}_1(t\gtrsim3\tau) = -\frac{\gamma^2 \tau \alpha \sqrt{1-\alpha^2}}{4\omega} \sqrt{\frac{\pi}{2}} \Biggl[&
    \Bigl(1+\frac{\omega_1}{\omega_2}\Bigr)
    +e^{-\frac{\tau^2\omega_r^2}{2}}\Bigl(1-\frac{\omega_1}{\omega_2}\Bigr)\Biggr]\cos\omega_rt 
    \label{supp:phi1_smooth_gauss}
\end{aligned}
\end{equation}
\begin{equation}
\begin{aligned}
    \bar{\theta}_1(t\gtrsim3\tau) = \frac{\gamma^2 \tau \alpha \sqrt{1-\alpha^2}}{4\omega} \sqrt{\frac{\pi \omega_1}{2 \omega_2}}
    \Biggl[\Bigl(1+\frac{\omega_2}{\omega_1}\Bigr)-e^{-\frac{\tau^2\omega_r^2}{2}}\Bigl(1-\frac{\omega_2}{\omega_1}\Bigr)\Biggr]\sin\omega_rt
    \label{supp:theta1_smooth_gauss}
\end{aligned}
\end{equation}

Taking into account only the dominant term, we can get an effective field of the inverse Faraday effect. Using the result from Section \ref{appendix2} one can obtain effective field of inverse Faraday effect for gaussian pulse in the case $\tau\gg\frac{2\pi}{\omega}$:
\begin{equation}
    H_{\text{IFE}}^{m}=-\frac{\gamma}{2\omega i}[\textbf{h}_0\times\textbf{h}_0^*]_z f^2(t)
    \label{supp:hife_classical}
\end{equation}
The details and definitions of the effective magnetic field are analyzed in the following section.

\section{Effective magnetic field}\label{appendix2}
This section explains how the effective magnetic field $\textbf{H}_{\text{eff}}$ is determined. We define the effective field as a slow function proportional to $f^2(t)$, each component of which has an envelope character, rather than a highly oscillating component.

In this section there is only one contribution to Zeeman energy caused by slow $\textbf{H}_{\text{eff}}(t) = (H_x(t), H_y(t), H_x(t))$ (written in Cartesian coordinates), time dependence is non oscillating functions. Thus, $V_h = - M (H_x(t)\sin\theta\cos\phi+H_y(t)\sin\theta\sin\phi+H_z(t)\cos\theta)$. In this configuration, equilibrium state of the magnetization is determined by $\phi_0 = \pi/2, 3\pi/2$ and $\theta_0 = \pi/2, 3\pi/2$. Using equations \ref{supp:phi_dot} and \ref{supp:theta_dot} and decomposing to the first order by $\theta_1, \phi_1 \ll 1$ we have for biaxial anisotropy or uniaxial anisotropy under external magnetic field along $y$ axis:
\begin{align}
\dot{\phi}_1 -\omega_1\theta_1=& \gamma (H_y\theta_1+H_z)\approx\gamma H_z,\label{phi_dot_hife}\\
\dot{\theta}_1 + \omega_2\phi_1=& -\gamma(H_x+H_y\phi_1)\approx -\gamma H_x.\label{theta_dot_hife}
\end{align}
The exact solution for above equations \ref{phi_dot_hife} and \ref{theta_dot_hife} follows
\begin{align}
{\phi}_1 =& \gamma \int\limits_{-\infty}^t d\tilde{t} \biggr[H_{z}(\tilde{t})\cos(\omega_r(t-\tilde{t})) (\tilde{t})-H_x(\tilde{t})\sqrt{\frac{\omega_1}{\omega_2}}\sin(\omega_r(t-\tilde{t}))\biggl],\label{supp:phi_hife}\\
{\theta}_1 =& -\gamma \int\limits_{-\infty}^t d\tilde{t} \biggr[H_{z}(\tilde{t})\sqrt{\frac{\omega_2}{\omega_1}}\sin(\omega_r(t-\tilde{t})) +H_{x}(\tilde{t})\cos(\omega_r(t-\tilde{t}))\biggl].\label{supp:theta_hife}
\end{align}

\begin{figure}[h]
    \centering
    \includegraphics[width=1\linewidth]{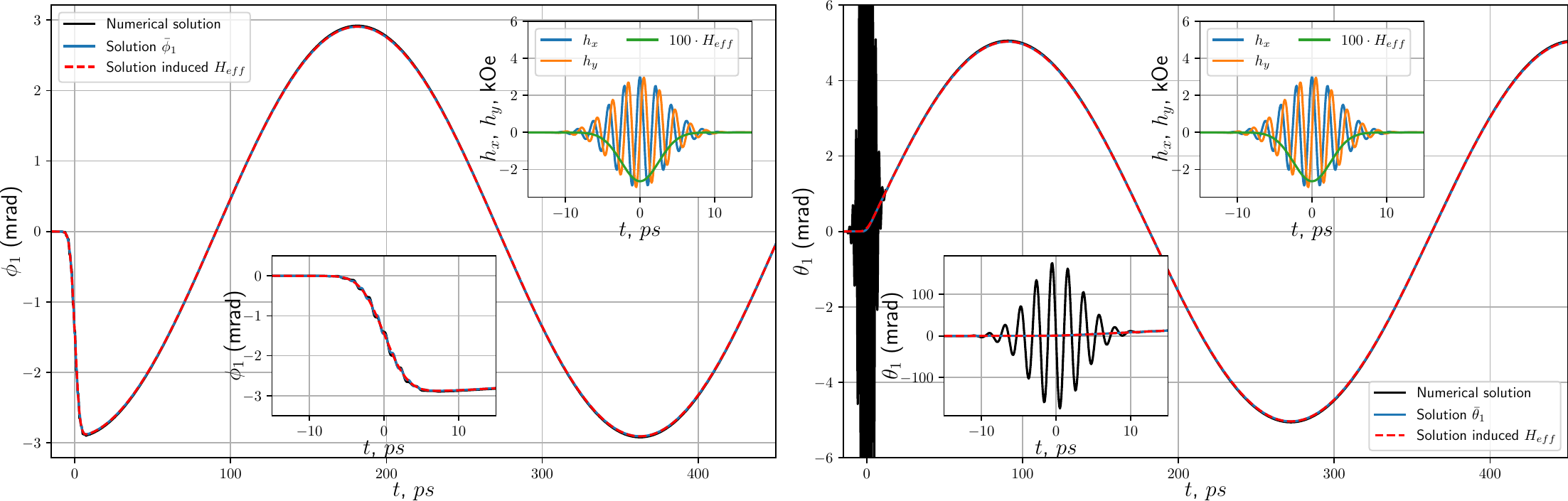}
    \caption{\justifying{
    The numerical solution to linearized equations \ref{phi1_dot} and \ref{theta1_dot} for pulse with circular polarization with $\tau = 5$ps and $\omega=3$THz are presented using black solid line. Constructed smooth oscillating part of solution (Equations \ref{supp:phi1_smooth_gauss} and \ref{supp:theta1_smooth_gauss}) are shown using blue solid line. Numerical solution of differential equations \ref{phi_dot_hife} and \ref{theta_dot_hife}, where effective magnetic field is presented in Equation \ref{supp:eff_long}, is illustrated by red dashed line.
    The parameters used in the numerical simulation of the linearized equations are selected as \(\gamma = 1.76 \times 10^{-5} \, \frac{1}{\mathrm{ps} \cdot \mathrm{Oe}}\), \(\omega_1 = 10\, \mathrm{GHz}\), \(\omega_2 = 30\, \mathrm{GHz}\), \(h_0 = 100\, \mathrm{Oe}\), and \(\varphi = 0\). }}
    \label{supp:figure1}
\end{figure}
\begin{figure}[h]
    \centering
    \includegraphics[width=1\linewidth]{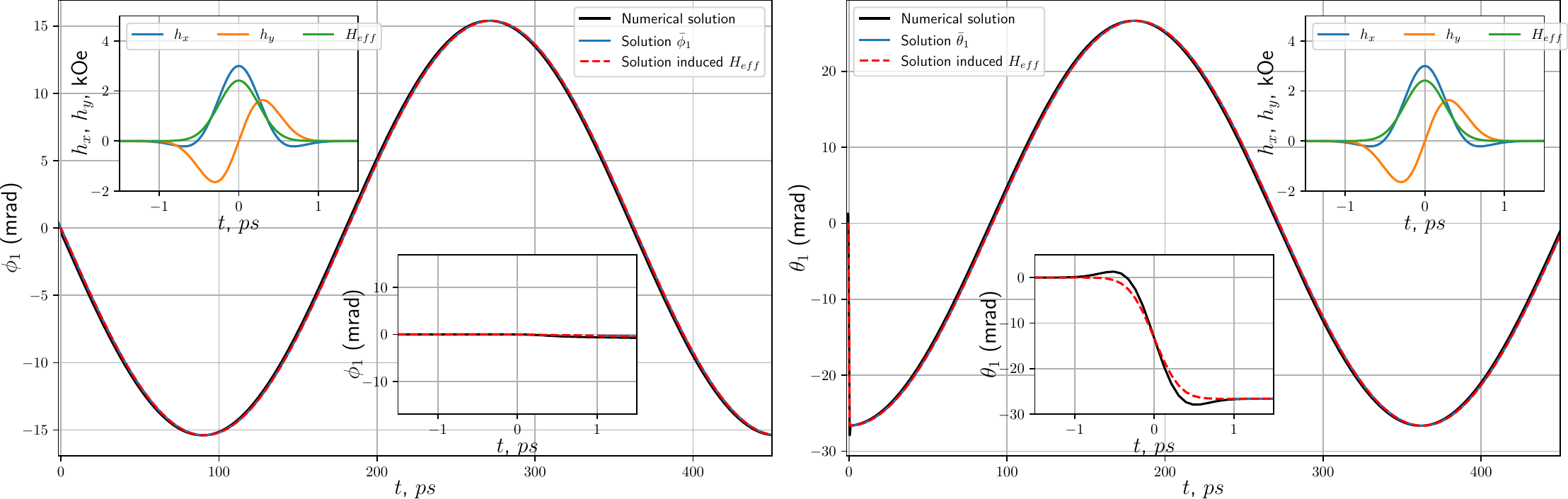}
    \caption{\justifying{
    The numerical solution to linearized equations \ref{phi1_dot} and \ref{theta1_dot} for pulse with circular polarization with $\tau = 0.7$ps and $\omega=3$THz are presented using black solid line. Constructed smooth oscillating part of solution (Equations \ref{supp:phi1_smooth_gauss} and \ref{supp:theta1_smooth_gauss}) are shown using blue solid line with $t>1ps$, since the solution is correct in the case $t\gtrsim 1 ps$. Numerical solution of differential equations \ref{phi_dot_hife} and \ref{theta_dot_hife}, where effective magnetic field is presented in Equation \ref{supp:eff_long}, is illustrated by red dashed line.
    The parameters used in the numerical simulation of the linearized equations are selected as \(\gamma = 1.76 \times 10^{-5} \, \frac{1}{\mathrm{ps} \cdot \mathrm{Oe}}\), \(\omega_1 = 10\, \mathrm{GHz}\), \(\omega_2 = 30\, \mathrm{GHz}\), \(h_0 = 100\, \mathrm{Oe}\), and \(\varphi = 0\). }}
    \label{supp:figure2}
\end{figure}

In the limit of long pulses the solution induced by effective magnetic field is consistent with the the linearized equations of motion (Equations \ref{supp:phi1_dot} and \ref{supp:theta1_dot}) and smooth oscillating part (Equations \ref{supp:phi1_smooth_gauss} and \ref{supp:theta1_smooth_gauss}). We get the effective magnetic field in the limit of long pulses $\tau\omega\gg2\pi$
\begin{equation}
    \mathbf{H}_{eff} =  -\frac{\gamma}{2\omega i}[\textbf{h}_0\times\textbf{h}_0^*]_z \, e^{-2\frac{t^2}{\tau^2}}\mathbf{z}.
    \label{supp:eff_long}
\end{equation}
The comparison of numerical solution of Eq.~A.1 and A.2 with linearized analytical solution Eq.~A.11 and A.12 and  numerical solution of spin dynamics induced by effective magnetic field Eq.~\ref{supp:phi_hife} and \ref{supp:theta_hife} is presented in the Figure \ref{supp:figure1}.

In the case of $\tau\lesssim 2\pi/\omega$ we get from decomposition by a small parameter near zero and summation
\begin{align}
\phi_1(t \gtrsim 3\tau)&= \gamma h_0\tau\cos\varphi \sqrt{\frac{\pi\omega_1}{2\omega_2}}e^{-\tau^2\omega^2/4}\sin\omega_rt 
    \label{supp:phi_faraday_gauss_short},\\
\theta_1(t \gtrsim 3\tau)&=\gamma h_0\tau\cos\varphi \sqrt{\frac{\pi}{2}}e^{-\tau^2\omega^2/4}\cos\omega_rt.
    \label{supp:theta_faraday_gauss_short}
\end{align}
Therefore, after comparing Eq.~\ref{supp:phi_faraday_gauss_short} and \ref{supp:theta_faraday_gauss_short} with Eq.~\ref{supp:phi_hife} and \ref{supp:theta_hife} we have the effective magnetic field as
\begin{equation}
    \mathbf{H}_{eff} = \sqrt{2}\, h_0 e^{-\frac{\tau^2\omega^2}{4}}f^2(t)\,\mathbf{x} .
\end{equation}
In the limit of short pulses the solution induced by effective magnetic field is consistent with the the linearized equations of motion (Equations \ref{supp:phi1_dot} and \ref{supp:theta1_dot}) and smooth oscillating part (Equations \ref{supp:phi_faraday_gauss_short} and \ref{supp:theta_faraday_gauss_short}). The comparison of time dependence os these values are presented in the Figure \ref{supp:figure2} 

\section{Fast Fourier transform spectrа of the excited spin dynamics}\label{appendix_spectrum}
We define the Fourier transform of a function $f(t)$ by $\mathcal{F}\{f\}(\Omega)
  \;=\;
  \int_{-\infty}^{\infty} f(t)\,e^{-i\,\Omega t}\,dt,$
where $\Omega\in \mathbb{R}$ denotes the angular frequency in rad/s. Let
$\tau>0$ be the temporal scale (width) of the Gaussian, $\omega>0$ be the carrier angular frequency.
The Fourier transforms of the pulse characteristic function $f_{cos}=\cos(\omega t)\,e^{-t^2/\tau^2}$ and $f_{sin}=\sin(\omega t)\,e^{-t^2/\tau^2}$ can be written as sums of two shifted Gaussians:
\begin{align}
\mathcal{F}\{f_{cos}\}(\Omega) = \sqrt{\pi}\,\tau\,
  \exp\!\left[-\frac{\tau^2}{4}\left(\Omega^2+\omega^2\right)\right]\,
  & \cosh\!\left(\frac{\tau^2}{2}\,\Omega\omega\right)=\\
  \frac{\sqrt{\pi}\,\tau}{2} & \left[
    \exp\!\left(-\frac{(\Omega-\omega)^2\tau^2}{4}\right)
    +
    \exp\!\left(-\frac{(\Omega+\omega)^2\tau^2}{4}\right)
  \right],\nonumber
  \\
\mathcal{F}\{f_{sin}\}(\Omega) = -\,i\,\sqrt{\pi}\,\tau\,
  \exp\!\left[-\frac{\tau^2}{4}\left(\Omega^2+\omega^2\right)\right]\,
  &\sinh\!\left(\frac{\tau^2}{2}\,\Omega\omega\right) = \\
  \frac{\sqrt{\pi}\,\tau}{2i} &\left[
    \exp\!\left(-\frac{(\Omega-\omega)^2\tau^2}{4}\right)
    -
    \exp\!\left(-\frac{(\Omega+\omega)^2\tau^2}{4}\right)
  \right] \nonumber
\end{align}
Therefore, a pulse of any polarization in the case of long pulses will have the form of a Gaussian in frequency with a peak at $\Omega = \omega$. In the case of short pulses FFT shows that $|\mathcal{F}\{h_x\}|, |\mathcal{F}\{h_y\}|\not=0$ at $\Omega=0$ (Fig.~\ref{fig:fft-side-by-side}).

\begin{figure*}[h]
  \centering
  \begin{subfigure}{0.49\textwidth}
    \centering
    \includegraphics[width=\linewidth]{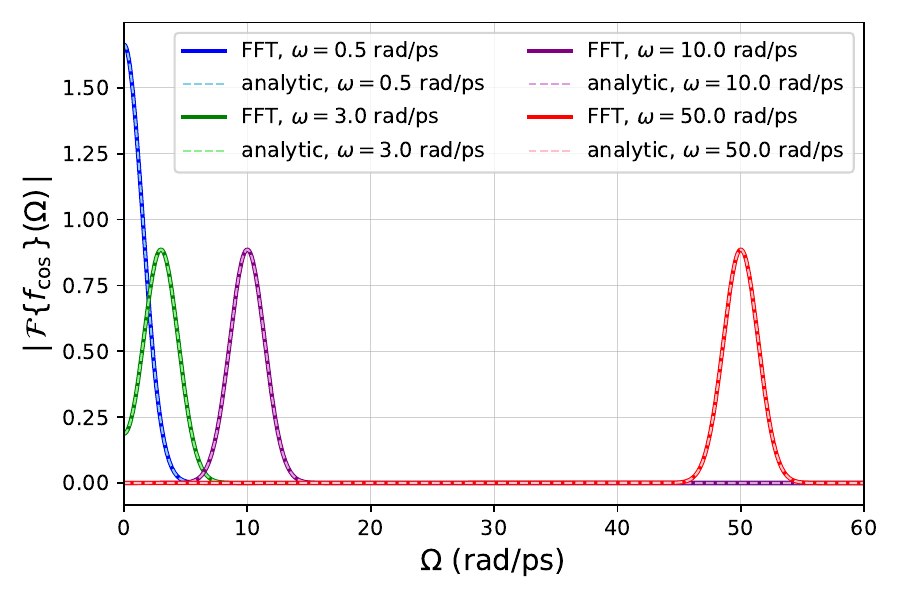}
    \label{fig:fft-cos}
  \end{subfigure}\hfill
  \begin{subfigure}{0.49\textwidth}
    \centering
    \includegraphics[width=\linewidth]{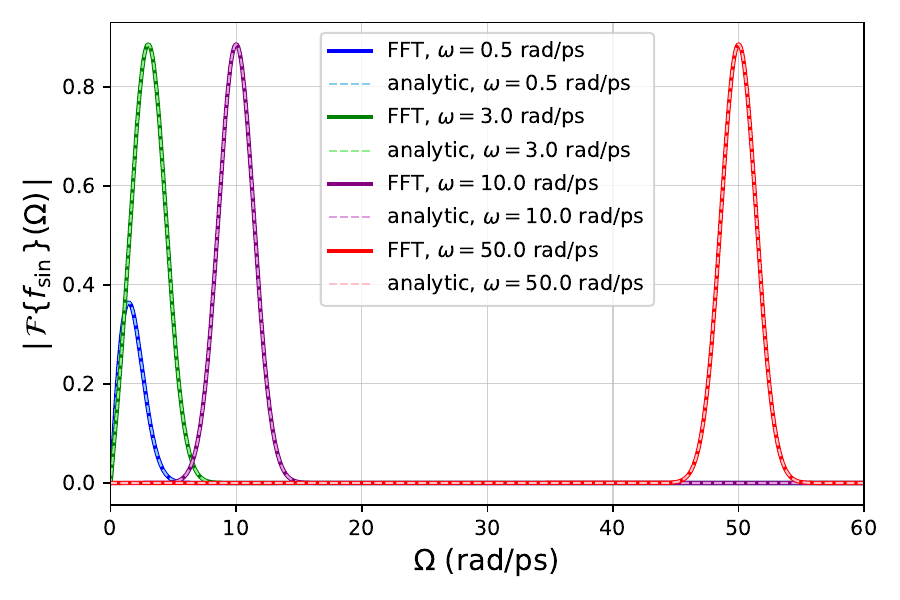}
    \label{fig:fft-sin}
  \end{subfigure}
  \caption{\justifying{The absolute value of $\mathcal{F}\{f_{cos}\}(\Omega)$ and $\mathcal{F}\{f_{sin}\}(\Omega)$  for $\tau=1\,\mathrm{ps}$ and $\omega\in\{0.5,3,10,50\}\,\mathrm{rad/ps}$. The solid line corresponds to numerical fast Fourier transform (FFT), the dashed line for analytical formula. In this case the examples with $\omega\in\{0.5,3\}\,\mathrm{rad/ps}$ correspond to resonant excitation at $\omega_r=0.0173$rad/ps, and $\omega\in\{10,50\}\,\mathrm{rad/ps}$ correspond to non-resonant}}
  \label{fig:fft-side-by-side}
\end{figure*}

\begin{figure*}[h!]
  \centering
    \includegraphics[width=\linewidth]{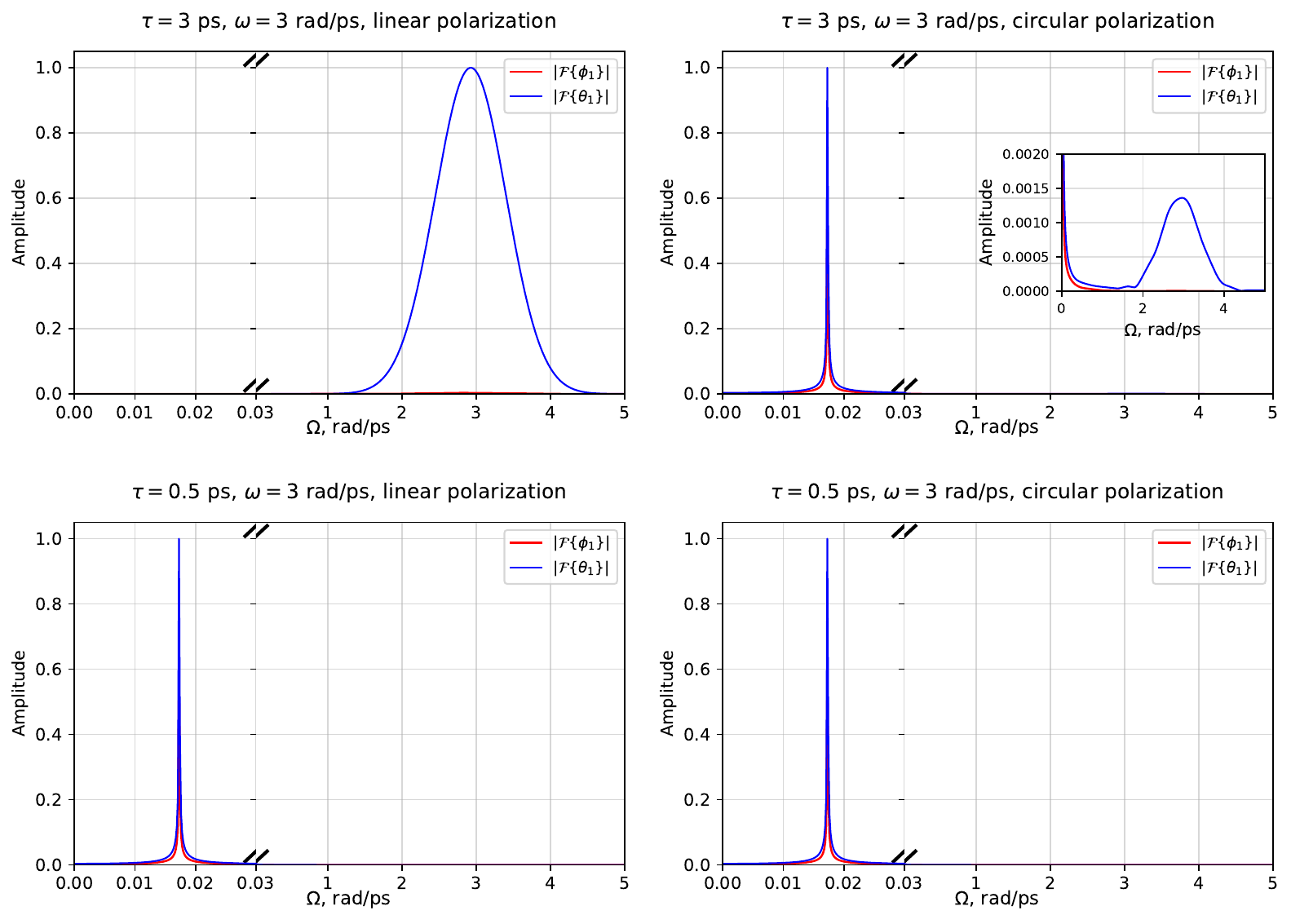}
  \caption{\justifying{The FFT absolute value of spin dynamics presented in Fig. \ref{fig:figure2} and \ref{fig:figure3}. }}
  \label{fig:fft_dynamics}
\end{figure*}

In our article we get the solutions of linearized equations presented in the figures \ref{fig:figure2} and \ref{fig:figure3}. The magnetization angles $\theta$ and $\phi$ are perturbed from the equilibrium positions by the magnetic component of electromagnetic pulse $\mathbf{h}$ (the fast Fourier transform (FFT) to $f_{sin}$ and $f_{cos}$ is shown in Fig. \ref{fig:fft-side-by-side}), the magnetization follows the pulse on frequency $\omega$. After the forced oscillations of magnetization due to the pulse, the magnetization precess around the equilibrium position on the resonant frequency $\omega_r$. Therefore, in the fast Fourier transform the peak values are observed in $\omega$ and $\omega_r$ (Fig. \ref{fig:fft_dynamics}).

We see that in the case of long pulses FFT shows no distinct frequency $\omega_r$ (Fig. \ref{fig:fft_dynamics}), in all other cases the resonant frequency is observed in FFT spectrum. In the case of short pulses no peak at angular frequency $\omega$ is observed, but in the case of long pulses the forced oscillations at frequency $\omega$ is presented.
Note, that spin dynamics is excited non-resonantly in the case of long pulses.

\section{Visible and near infrared range}\label{appendix4}
In laboratory conditions, femtosecond laser pulses in the visible range are commonly used due to their high energy concentrated in a single pulse. Considering a fluence of $1~\text{mJ}~\text{cm}^{-2}$, pulse duration $\tau = 250$~fs, and angular frequency $\omega=2\pi \cdot 500$~rad~ps$^{-1}$, which corresponds to a visible light wavelength of $\lambda=\frac{2\pi c}{\omega}= 600$~nm, we can calculate the electric field strength of such a pulse. For these parameters, the electric field amplitude reaches about $1.23~\text{MV/cm}$. This electric field intensity corresponds to a magnetic field amplitude $h_0 = 4100$~Oe (the numerical solution of Eq.~A.2 and A.3 is presented in Fig.~\ref{supp:figure5}). These laser pulse parameters are characteristic of relatively long pulses in the femtosecond regime, corresponding to $\tau\omega\gg2\pi$. This contribution to the dynamics of magnetization can be obtained, for example, by exciting the TE mode \cite{gribova2024inverse}. Thus, it will be possible to separate the electrodipole and magnetodipole contributions to the effective magnetic field of inverse Faraday effect. 

\begin{figure}[h]
    \centering
    \includegraphics[width=0.7\linewidth]{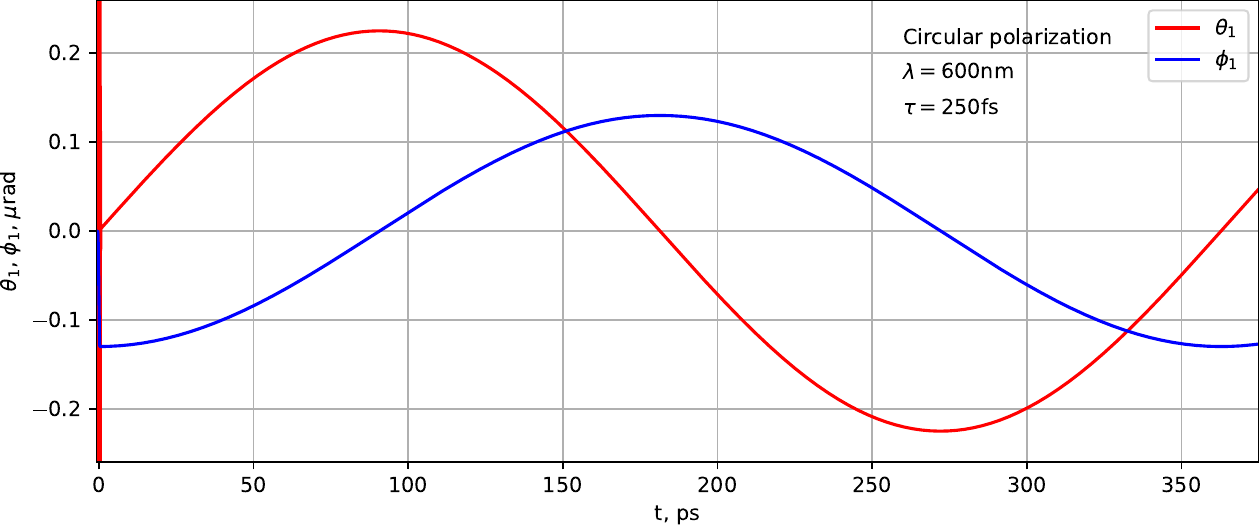}
    \caption{\justifying{
    The numerical solution to linearized equations \ref{phi1_dot} and \ref{theta1_dot} for pulse with circular polarization with $\tau = 250$~fs and $\omega=3142$~rad/ps ($\lambda=600$~nm) are presented using blue and red solid lines, respectively.
    The parameters used in the numerical simulation of the linearized equations are selected as \(\gamma = 1.76 \times 10^{-5} \, \frac{1}{\mathrm{ps} \cdot \mathrm{Oe}}\), \(\omega_1 = 10\, \mathrm{GHz}\), \(\omega_2 = 30\, \mathrm{GHz}\), \(h_0 = 4100\, \mathrm{Oe}\), and \(\varphi = 0\), $\psi=0$, $\alpha=1/\sqrt{2}$. }}
    \label{supp:figure5}
\end{figure}

\clearpage
\end{widetext}
\end{document}